\documentclass[lettersize,journal]{IEEEtran}
\UseRawInputEncoding
\usepackage{longtable}
\usepackage{amsmath,amsfonts}
\usepackage{algorithmic}
\usepackage{algorithm}
\usepackage{array}
\usepackage[caption=false,font=normalsize,labelfont=sf,textfont=sf]{subfig}
\usepackage{xurl}
\usepackage{url}
\usepackage{textcomp}
\usepackage{stfloats}
\usepackage{url}
\usepackage{enumitem} 
\usepackage{verbatim}
\usepackage{graphicx}
\usepackage{xcolor}
\usepackage{cite}
\usepackage{acro}
\usepackage{tabularx,booktabs,textcomp}
\usepackage{bm}
\usepackage{tabularx}
\usepackage[justification=centering]{caption}
\usepackage{fancyhdr}
\usepackage{amsthm}

\usepackage{amsmath,amssymb,amsthm}

\usepackage{pdflscape}

\usepackage{acro}
\DeclareAcronym{EV}{
  short=EV,
  long=Electric Vehicle,
}
\DeclareAcronym{ML}{
  short=ML,
  long=Machine Learning,
}
\DeclareAcronym{EVCI}{
  short=EVCI,
  long=Electric Vehicle Charging Infrastructure,
}
\DeclareAcronym{P2P}{
  short=P2P,
  long=Peer-to-Peer,
}
\DeclareAcronym{ESS}{
  short=ESS,
  long=Energy Storage Systems,
}

\DeclareAcronym{DER}{
  short=DER,
  long=Distributed Energy Resources,
}

\DeclareAcronym{RL}{
  short=RL,
  long=Reinforcement Learning,
}

\DeclareAcronym{V2G}{
  short=V2G,
  long=Vehicle to Grid,
}

\DeclareAcronym{DRL}{
  short=DRL,
  long=Deep Reinforcement Learning,
}

\DeclareAcronym{DR}{
  short=DR,
  long=Demand Response,
}

\DeclareAcronym{ToU}{
  short=ToU,
  long=Time of Use,
}
\DeclareAcronym{EVCS}{
  short=EVCS,
  long=Electrical Vehicle Charging Station,
}

\DeclareAcronym{PV}{
  short=PV,
  long=Photovoltaic,
}

\DeclareAcronym{CHP}{
  short=CHP,
  long=Combined Heat and Power,
}

\DeclareAcronym{PL}{
  short=PL,
  long=Parking Lots,
}
\DeclareAcronym{SoC}{
  short=SoC,
  long=State of Charge,
}

\DeclareAcronym{SoD}{
  short=SoD,
  long=State of Discharge,
}
\DeclareAcronym{WM}{
  short=WM,
  long=Wholesale Market,
}

\DeclareAcronym{G2V}{
  short=G2V,
  long=Grid to Vehicle,
}

\DeclareAcronym{V2H}{
  short=V2H,
  long=Vehicle to Home,
}

\DeclareAcronym{SAEV}{
  short=SAEV,
  long=Shared Autonomous Electric Vehicle,
}

\DeclareAcronym{HEMS}{
  short=HEMS,
  long=Home Energy Management Systems,
}

\DeclareAcronym{V2V}{
  short=V2V,
  long=Vehicle to Vehicle,
}

\DeclareAcronym{PSO}{
  short=PSO,
  long=Particle Swarm Optimization,
}
\DeclareAcronym{EDS}{
  short=EDS,
  long=Electrical Distribution Systems,
}
\DeclareAcronym{PPAs}{
  short=PPAs,
  long=Power Purchase Agreements,
}
\DeclareAcronym{FiTs}{
  short=FiTs,
  long=Feed-in-Tariffs,
}
\DeclareAcronym{FDI}{
  short=FDI,
  long=Fault Data Injection,
}

\DeclareAcronym{DoS}{
  short=DoS,
  long=Denial-of-Service,
}
\DeclareAcronym{SG}{
  short=SG,
  long=Smart Grid,
}
\DeclareAcronym{EMS}{
  short=EMS,
  long=Energy Management Systems,
}
\DeclareAcronym{MGs}{
  short=MGs,
  long=Microgrids,
}
\DeclareAcronym{PEV}{
  short=PEV,
  long=Plug-in Electric Vehicle,
}
\DeclareAcronym{PHEV}{
  short=PHEV,
  long=Plug-in Hybrid Electric Vehicle,
}

\DeclareAcronym{MDP}{
  short=MDP,
  long=Markov Decision Process,
}

\DeclareAcronym{SVM}{
  short=SVM,
  long=Support Vector Machine,
}

\DeclareAcronym{DDoS}{
  short=DDoS,
  long=Distributed Denial of Services,
}

\DeclareAcronym{RC}{
  short=RC,
  long=Repair Crew,
}

\DeclareAcronym{IEA}{
  short=IEA,
  long=International Energy Agency,
}
\DeclareAcronym{GHG}{
  short=GHG,
  long=Green House Gases,
}

\DeclareAcronym{VPP}{
  short=VPP,
  long=Virtual Power Plant,
}
\begin{document}

\title{Energy Sharing among Resources within Electrical Distribution Systems: A Systematic Review}

\author{{G Hari Krishna,  K. Victor Sam Moses Babu, Divyanshi Dwivedi, Pratyush Chakraborty, Pradeep Kumar Yemula, Mayukha Pal$^{*}$}
\thanks{(Corresponding author: $^{*}$Mayukha Pal)}
\thanks{Mr. G Hari Krishna  is a Data Science Research Intern at ABB Ability Innovation Center, Hyderabad 500084, India and also a Research Scholar at the Department of Electrical and Electronics Engineering, BITS Pilani Hyderabad Campus, Hyderabad 500078, IN.}
\thanks{Mr. K. Victor Sam Moses Babu is a Data Science Research Intern at ABB Ability Innovation Center, Hyderabad 500084, India and also a Research Scholar at the Department of Electrical and Electronics Engineering, BITS Pilani Hyderabad Campus, Hyderabad 500078, IN.}
\thanks{Mrs. Divyanshi Dwivedi is a Data Science Research Intern at ABB Ability Innovation Center, Hyderabad 500084, India, and also a Research Scholar at the Department of Electrical Engineering, Indian Institute of Technology, Hyderabad 502205, IN.}
\thanks{Dr. Pratyush Chakraborty is an Assistant Professor with the Department of Electrical and Electronics Engineering, BITS Pilani Hyderabad Campus, Hyderabad 500078, IN.}
\thanks{Dr. Pradeep Kumar Yemula is an Associate Professor with the Department of Electrical Engineering, Indian Institute of Technology, Hyderabad 502205, IN.}
\thanks{Dr. Mayukha Pal is with ABB Ability Innovation Center, Hyderabad-500084, IN, working as Global R\&D Leader – Cloud \& Analytics (e-mail: mayukha.pal@in.abb.com).}
}

\maketitle
\begin{abstract}
The rapid increase in \ac{EV} adoption provides a promising solution for reducing carbon emissions and fossil fuel dependency in transportation systems. However, the increasing numbers of \ac{EV}s pose significant challenges to the electrical grids. In addition, the number of \ac{DER} and \ac{MGs} is increasing on a global scale to meet the energy demand, consequently changing the energy infrastructure. Recently, energy-sharing methods have been proposed to share excess energy from \ac{DER}s and \ac{EV}s in \ac{EVCI} and \ac{MGs}. Accommodating this sharing mechanism with the existing electrical distribution systems is a critical issue concerning the economic, reliability, and resilience aspects. This study examines the ever-changing field of EVCI and the critical role of \ac{P2P} energy trading in mitigating the problems with grid management that result from unorganized EV charging and intermittency in \ac{DER}. Also, the possibilities of energy sharing in electrical distribution systems for microgrids and EVCI on various energy-sharing methods and algorithms are discussed in detail. Furthermore, the application of market clearing algorithms like game theory, double auction theory, blockchain technology, optimization techniques, machine learning algorithms, and other models from the existing literature are presented. This paper discusses the policies, economic benefits, environmental impacts, societal advantages, and challenges in distribution systems related to sharing in \ac{EVCI} and \ac{MGs}. A roadmap for future research and sharing strategies is provided to guide policymakers, researchers, and industry stakeholders toward a sustainable, resilient, and efficient energy market by integrating P2P technology into EVCIs and \ac{MGs}.
\end{abstract}

   
\printacronyms

\section{Introduction}
\label{section:Intro}

\subsection{Background \& Motivation}
\begin{figure*}
    \centering
    \includegraphics[width=0.9\linewidth]{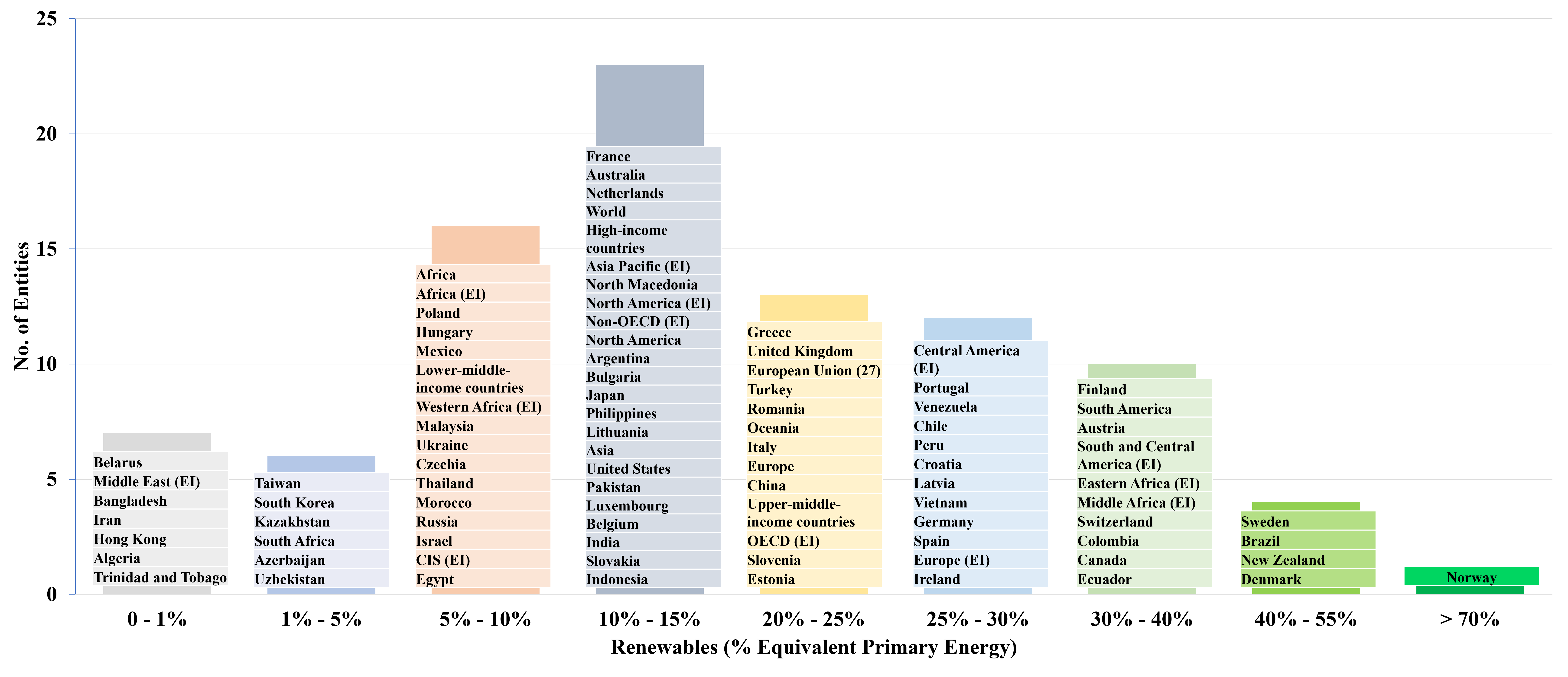}
    \caption{Renewables (\% equivalent primary energy) for entities across the world in 2022
    \cite{owid-renewable-energy}.}
    \label{fig:2}   
\end{figure*}

As the world is looking to achieve sustainable development goals \cite{UN_SDG}, the realization of optimal utilization of resources has increased. The usage of clean energy, reduction of greenhouse gases, and energy-efficient systems is gaining importance. The Paris Agreement (2015), in alignment with the Kyoto Protocol (1997) and endorsed by the majority of nations as confirmed in the United Nations Framework Convention on Climate Change (UNFCC), commits to the reduction of greenhouse gas (GHG) emissions and carbon footprint \cite{leggett2020united, fccc2015fccc}. This commitment includes the adoption of renewable energy sources, particularly solar energy, for electricity generation. As of 2022, renewable energy sources accounted for 14.2\% of the global energy \cite{owid-renewable-energy}. Fig. \ref{fig:2} shows the percentage breakdown of various countries' electricity generation through renewable sources. According to the \ac{IEA} \cite{EMR}, this percentage is expected to rise to 35\% by 2025, indicating a significant shift towards renewable energy on a global scale.

\begin{figure}[t]
    \centering
    \includegraphics[width=0.9\linewidth]{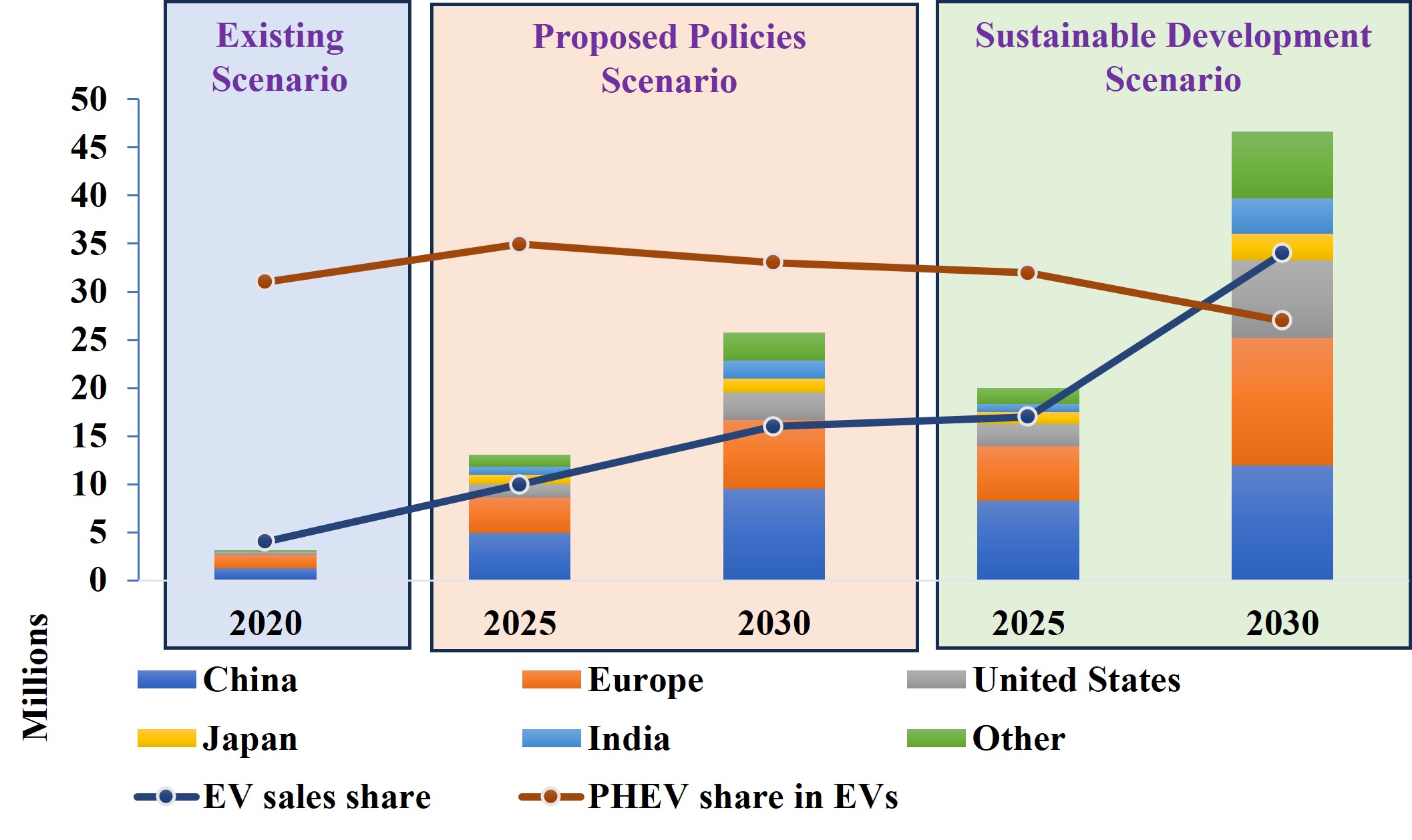}
    \caption{Projected EV sales with existing scenario, proposed policies scenario, and sustainable development scenario for the corresponding years from 2020 to 2030 \cite{index1}.}
    \label{fig:EVsales}
\end{figure}

The proliferation of electric vehicles is also gaining popularity as an alternative to conventional internal combustion engines and \ac{PHEV}. The world is seeing phenomenal changes in the transportation sector due to the surge in \ac{EV}s. The sale of \ac{EV}s touched a milestone of 10 million in the year 2020 with the existing policies implemented by the nations. With this rapid adaption rate, the \ac{IEA} predicts that the \ac{EV}s with the proposed policies are estimated to increase to 145 million by the year 2030 with a sustainable development scenario and expected share of 34\%, as illustrated in Fig. \ref{fig:EVsales} \cite{index1}.

\begin{figure}
    \centering
    \includegraphics[width=3.3in]{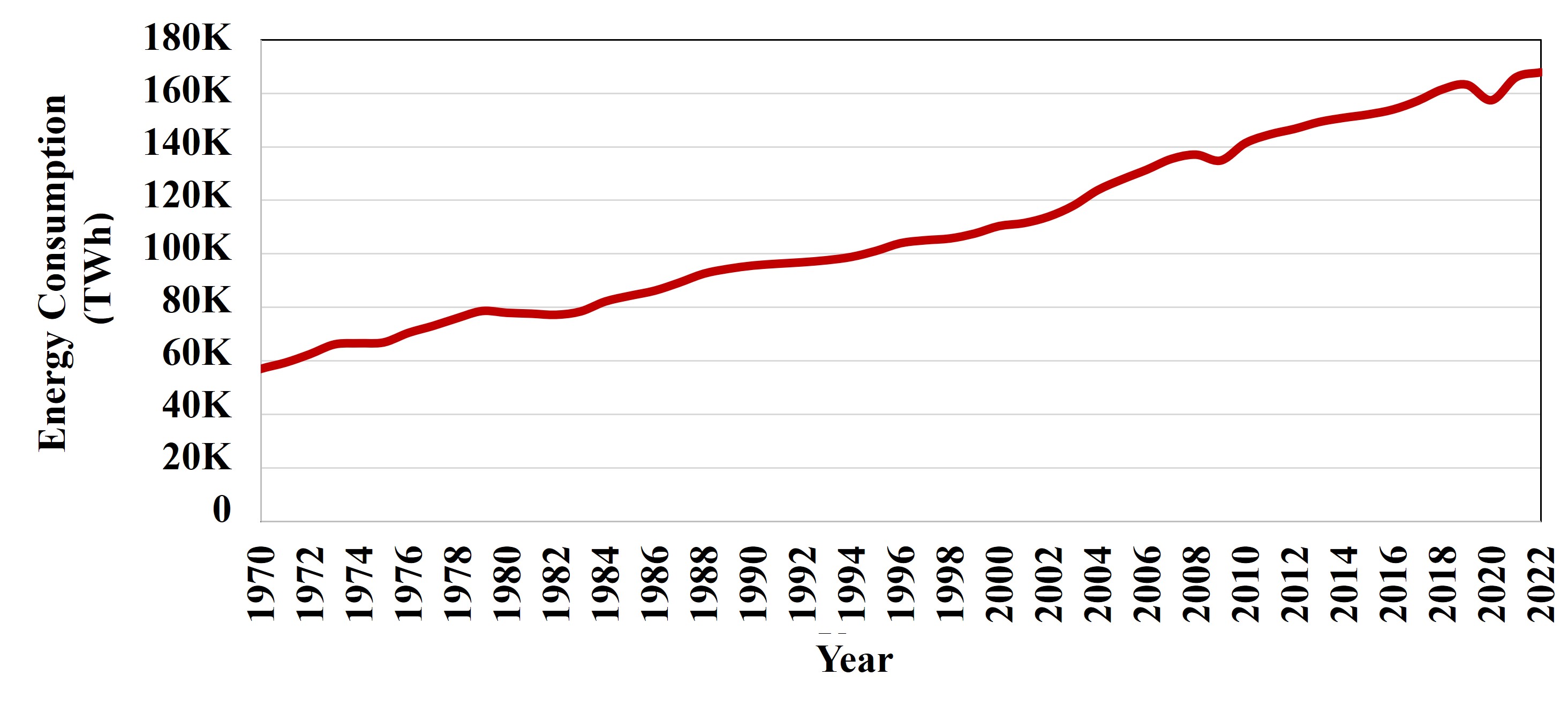}
    \caption{Primary energy consumption: World statistics\cite{owid-energy-production-consumption}.}
    \label{fig:1965-2022}
\end{figure}

The ever-increasing demand for energy presents a significant challenge to the current power grid infrastructure. The global energy consumption from 1970 to 2022, as depicted in Fig. \ref{fig:1965-2022}, exhibits a consistent upward trend, indicative of a pattern expected to increase further in the forthcoming years. In addition, the number of EVs on the road is increasing rapidly, leading to a higher demand for electricity. As a result, there is an urgent need to adopt \ac{DER}, enhance microgrid capabilities, and improve \ac{EVCI}.

\begin{figure*}
    \centering
    \includegraphics[width=0.7\linewidth]{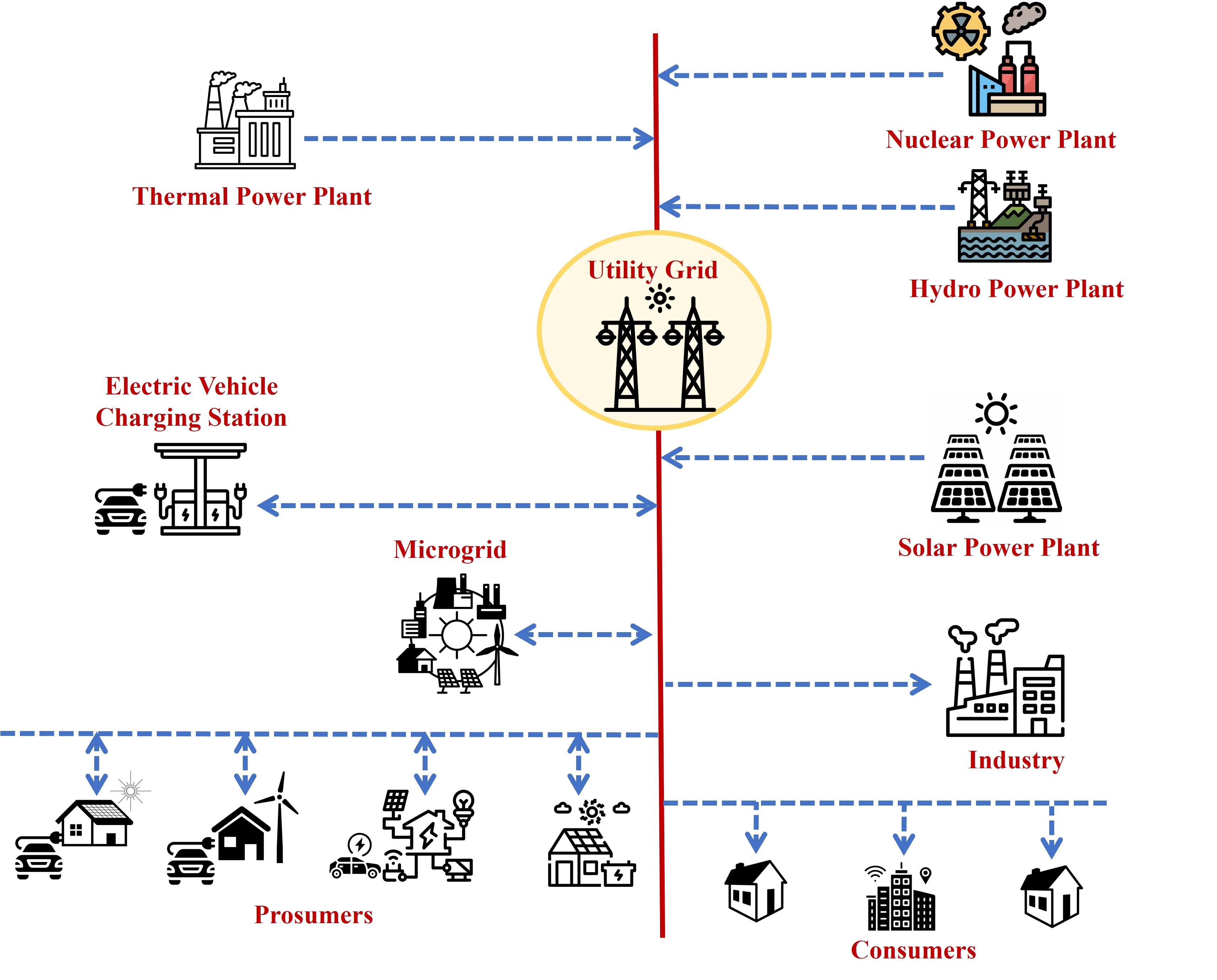}
    \caption{A visual representation of energy sharing.}
    \label{fig:ES}
\end{figure*}

The intersection of these three components, namely reduction of \ac{GHG}, adoption of \ac{EV}s, and increasing energy demand, resulted in the evolution of the concept called ``Energy sharing". The visual representation of energy sharing is shown in Fig. \ref{fig:ES}; it consists of various components such as microgrids, solar power plants, wind power plants, distributed energy sources, prosumers, and \ac{EVCS} that are integrated with the existing power grid. All these components are distributed throughout the power grid, sharing energy with the utility grid. This minimizes the gap between generation and demand, making the grid resilient. 

\subsection{Energy Sharing Trade-offs}
The primary driving factors for implementing energy sharing include:
\begin{enumerate}
    \item Reducing the carbon footprint from an environmental perspective by adapting \ac{DER} and EVs, and
    \item Meeting the increasing energy demand, thereby enhancing energy security.
\end{enumerate}

Furthermore, the deployment of \ac{EV}s and the implementation of vehicle fuel efficiency guidelines are pivotal assumptions for the upcoming decade. These factors are crucial in the context of introducing new vehicles into the fleet, developing charging infrastructure, and accelerating the decarbonization of the power grid. For sustainable development, microgrids and \ac{EVCI} serve as key enablers. Bimenyimana et al. \cite{bimenyimana2021integration} investigated the above problem by integrating solar \ac{PV} generation in microgrids to support \ac{EV} in Rwanda. Meanwhile, Wei Wu et al. \cite{wu2021benefits} revealed the benefits of integrating EVs into the power grid by conducting a pilot study. 
The development and deployment of microgrids where there is existing grid infrastructure is influenced by three main categories of factors: a combination of economic benefits, clean energy, and energy security is provided in \cite{hirsch2018microgrids}. Abdulgader Alsharif  et al.\cite{alsharif2021comprehensive} have provided a comprehensive overview of performance improvement and effective management of \ac{V2G} related to renewable energy resources through \ac{EMS}.

 
This paper presents an overview of energy sharing between microgrids, \ac{DER}, and \ac{EVCI} in electrical distribution systems. The main contributions are:

\begin{enumerate}
    \item It offers a comprehensive discussion covering various aspects of energy-sharing architecture and sharing models with case studies. 
    \item The technical aspects of \ac{P2P} energy trading are explored, including algorithms and models used for electricity trading. 
    \item Promising research directions to expand current practices are presented.
\end{enumerate}

The structure of this review is as follows: Section 2 includes the aspects of energy-sharing basics, architecture in the physical layer and virtual layer, implementation, and various sharing models, including \ac{P2P} sharing along with pilot projects and case studies. Billing mechanisms and pricing policies are also discussed, whereas, in Section 3, the benefits of energy sharing are discussed. In section 4, the technical aspects of \ac{P2P} energy trading are explored, including algorithms and models used for trading. Section 5 outlines the challenges, opportunities, and future scope of energy sharing in the physical layer and virtual layer, especially \ac{P2P}. Section 6 concludes the paper.

\begin{figure}
    \centering
    \includegraphics[width=1\linewidth]{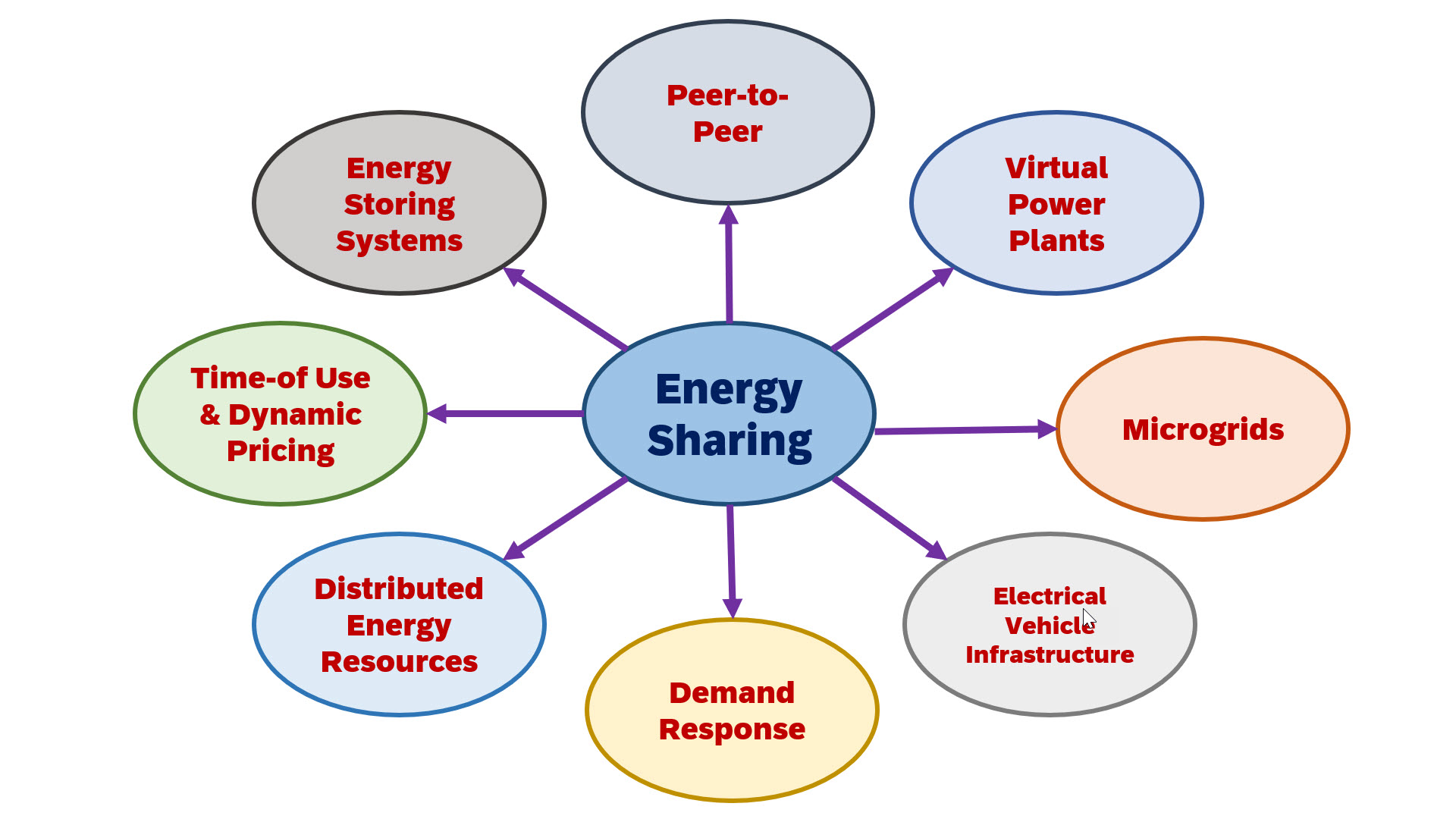}
    \caption{Types of Energy Sharing}
    \label{fig:EStypes}
\end{figure}

\section{Energy sharing architecture}
\begin{figure}
    \centering
    \includegraphics[width=1\linewidth]{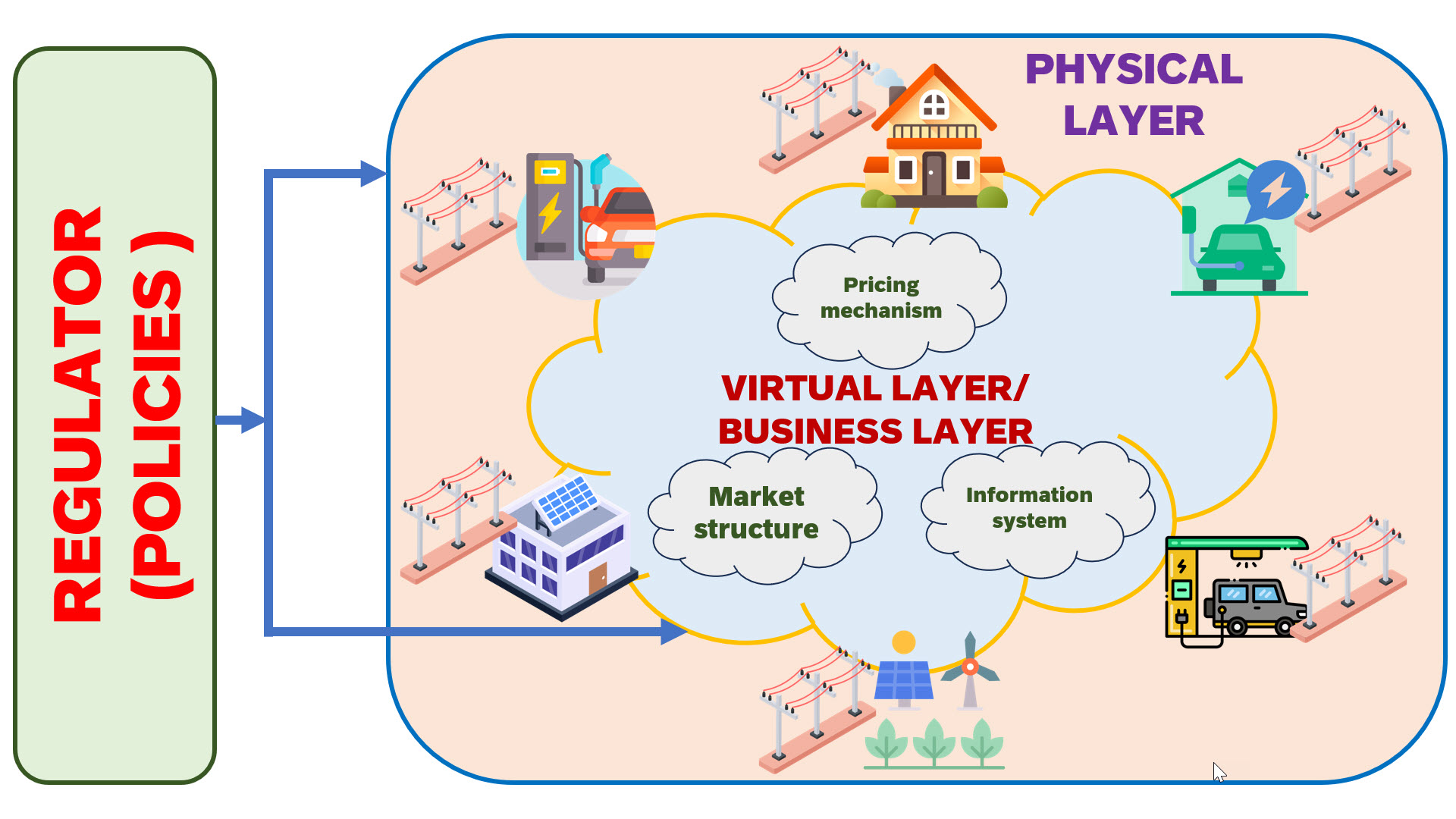}
    \caption{Energy Sharing Architecture}
    \label{fig:ES_arch}
\end{figure}

In modern smart grids and digital technologies, energy sharing is often divided into physical and virtual layers. These layers represent energy-sharing ecosystem components. Fig. \ref{fig:ES_arch} demonstrates different layers of the energy-sharing which has both physical layers and virtual layers along with a regulator.
\subsection{Physical Layer}
Energy-sharing infrastructure hardware is in the physical layer. This covers energy-generating, storing, and distributing equipment and facilities. Power plants, renewable energy sources (solar panels, wind turbines), energy storage systems (batteries), transmission lines, distribution networks, EV charging stations, and other physical assets comprise the physical layer.
\subsubsection{Energy Sharing Components-Physical Layer}
Fig. \ref{fig:EStypes} illustrates various methods for energy sharing in electrical distribution systems. The methods include \ac{P2P}, \ac{VPP}, \ac{DER}, microgrids, \ac{EVCI}, \ac{DR}, Time-of-use \& dynamic pricing, and \ac{ESS}. Governments globally are implementing policies to promote sustainable transportation and energy systems. These measures include incentives for electric vehicles, targets for renewable energy, and grid modernization. \cite{9775727}.

The concept of \ac{VPP} involves combining conventional dispatchable power plants, DERs, and ESS along with flexible loads to form a single controllable and manageable entity \cite{saboori2011virtual}. The article \cite{ghavidel2016review} discusses the components and modeling of a \ac{VPP}, its different types, such as the \ac{CHP} based VPP, \ac{DR}, and its participation in electricity markets. The article also covers the bidding strategies of VPPs. Meanwhile, \cite{yu2019uncertainties} presents how the optimization technique improves the running performance of VPPs when faced with uncertainties.

\ac{DR} involves consumers changing their electricity consumption patterns in exchange for incentives. Integrated \ac{DR} is a concept that involves coupling various forms of energy like electricity and heat. In distribution systems, this idea allows for the effective management of energy by responding to changes in demand. In their paper, Huang et al. (2019) \cite{huang2019demand} focus on the basic concept, design framework, and estimation of demand response utilizing software testing. To address uncertainties in multiple renewables and coordinate flexible \ac{DR} with \ac{EVCS}, Li et al. designed a new bi-level optimal dispatching model. Meanwhile, \cite{parrish2020systematic} examines the motivation, enablers, and barriers for consumer participation in demand response with residential demand response. In addition, \cite{lu2020fundamentals} discusses the fundamentals and business models for resource adequacy in the electricity market.

\ac{DER} are power generation sources mostly renewables like \ac{PV}, fuel cells, micro turbines, and \ac{ESS} that are directly connected to the electrical distribution systems. A detailed overview of \ac{DER} technologies and their benefits to the environment is given in \cite{akorede2010distributed} and in addition to that, coordination and controlling of \ac{DER} together will give rise to a concept called ``Microgrid". The global microgrid market surpassed 17 billion US dollars in 2022 and is expected to expand to 73.3 billion US dollars with a 22.4\% CAGR due to the growth of microgrids and \ac{DER}. The remarkable growth is primarily driven by the increasing demand for reliable, resilient, and off-grid power solutions.

Microgrid systems operates either in islanding or grid-connected mode. A comprehensive review of microgrid technologies is presented in \cite{jiayi2008review}. J. Driesen et al.\cite{driesen2008design} incorporated numerous \ac{DER} units into the \ac{EDS} to improve reliability and offer differentiated services, whereas state-of-art approaches, techniques, challenges in uncertainty modeling of \ac{DER} are conducted in \cite{zhang2019uncertainty}. Meanwhile, integration of \ac{DER} and its implications are studied by Lasseter et al.\cite{lasseter2002integration}. The adoption of \ac{DER} has significantly impacted the energy sector, particularly in microgrids. However, integrating these renewable sources leads to uncertainties due to their intermittent nature, which often causes voltage fluctuations and grid instability in electrical distribution systems.\cite{saeed2021review, jones2017renewable}.K. Victor Sam Moses Babu et al.\cite{10150022} analyzed the impact of \ac{P2P} sharing within microgrids and the results depicted the performance and system resilience improvement.

Behind-the-meter energy generation, \ac{EV}, batteries, inverters, and loads have shown unprecedented growth as small-scale \ac{DER} surges. It is projected that the number of households using rooftop solar photovoltaic systems will increase from 25 million in 2022 to 100 million by 2030 \cite{PV}. In addition, there is expected to be significant growth in the use of \ac{ESS} with capacity projected to increase from 20 GWh in 2020 to 160 GWh by 2026 \cite{batterystorage}.

\subsubsection{P2P Energy Sharing}
The increasing presence of prosumers, individuals, and entities that both produce and consume energy, has led to the decentralization and increased openness of the electrical network. Energy operators have evolved to offer services beyond just selling energy; they now provide the option for prosumers to rent transmission lines, enabling them to inject their surplus energy into the grid through net metering programs \cite{r1}. The implementation of energy-sharing and coordination strategies could lead to cost savings and reductions in greenhouse gas emissions. Utilizing excess renewable energy for electric vehicle charging, for instance, could bring about both economic benefits and ecological advantages. The advancement of smart grid technologies, IoT devices, and advanced control systems is empowering the implementation of more efficient and sophisticated energy sharing and coordination strategies \cite{dileep2020survey}.

Meeting our energy needs and advancing the integration of affordable, clean energy into the electrical grid are made possible in large part by these dispersed and varied energy sources. Nevertheless, there is a catch: the owners of these energy sources have to play two roles at once. These people have to buy and sell electricity in addition to fulfilling the roles of energy producers and consumers. This is where \ac{FiTs} programs come into play, providing a mechanism for these ``prosumers" to enter the electricity market and actively participate in it \cite{r4} and \cite{r5}.  Prosumers benefit from the \ac{FiTs} scheme, by receiving grid power in peak demand periods and selling surplus energy back to the grid when production exceeds their capacity. The downside is that they receive comparatively low compensation for this surplus electricity. In some countries, this financial restriction has led to the discontinuation or revision of these programmes.

As a response to these challenges, the emergence of \ac{P2P} electricity trading has gained prominence. It is the next level of smart grid energy management, and it lets people who make their own electricity trade it with their neighbors. P2P electricity trading represents an advanced stage in smart grid energy management, enabling individuals who generate their own electricity to engage in direct energy exchanges with their neighbors. With P2P trading, people who make extra energy could swap it within their community, which means they could earn more money. Within this setup, energy producers have the autonomy to define the operational dynamics and trading processes. This approach is fundamentally centered on empowering individuals to have greater control over their energy resources \cite{liu2019peer}.

\subsection{Virtual Layer}
The virtual layer of energy sharing replicates the important layer that utilizes the latest digital technologies for energy trading and it altered energy trading in the modern world. These technologies are transparent, easily accessible, and incredibly efficient. In the context of energy trading, the digital and software-based elements that make energy transaction management, execution, and optimization easier are referred to as the virtual layer. An ecosystem that is responsive and dynamic for energy trading is largely dependent on this layer. 

\subsubsection{Market Structures}
In a centralized energy market, the generation, transmission, and distribution of power are all under the direction and control of a single governing body or authority. In this model, decisions about energy dispatch, price, and grid management are often made by a central grid operator or government body. Centralized energy markets are frequently associated with dominating utility companies that own and operate most of the energy infrastructure.

A system in which multiple entities share decision-making, production, and distribution authority instead of a single central authority is known as a decentralized energy market. Numerous stakeholders participate in the management and operation of the energy system in a decentralized energy market, including towns, smaller utility companies, and independent power producers. More flexibility and adaptability are made possible by this concept, which typically refers to the integration of \ac{DER}. 

A decentralized, cooperative system of generating, distributing, and utilizing energy that involves community members actively creating and trading energy resources is known as a community-based energy market. According to this idea, local governments, cooperatives, and communities actively take part in the ownership, administration, and use of energy resources in their community. Community-based energy markets encourage communities to actively participate in the transition to sustainable and resilient energy systems. To establish a sustainable and locally controlled energy future, effective models involve community members, local authorities, energy experts, and other stakeholders.

\begin{figure*}
    \centering
    \includegraphics[width=0.8\linewidth]{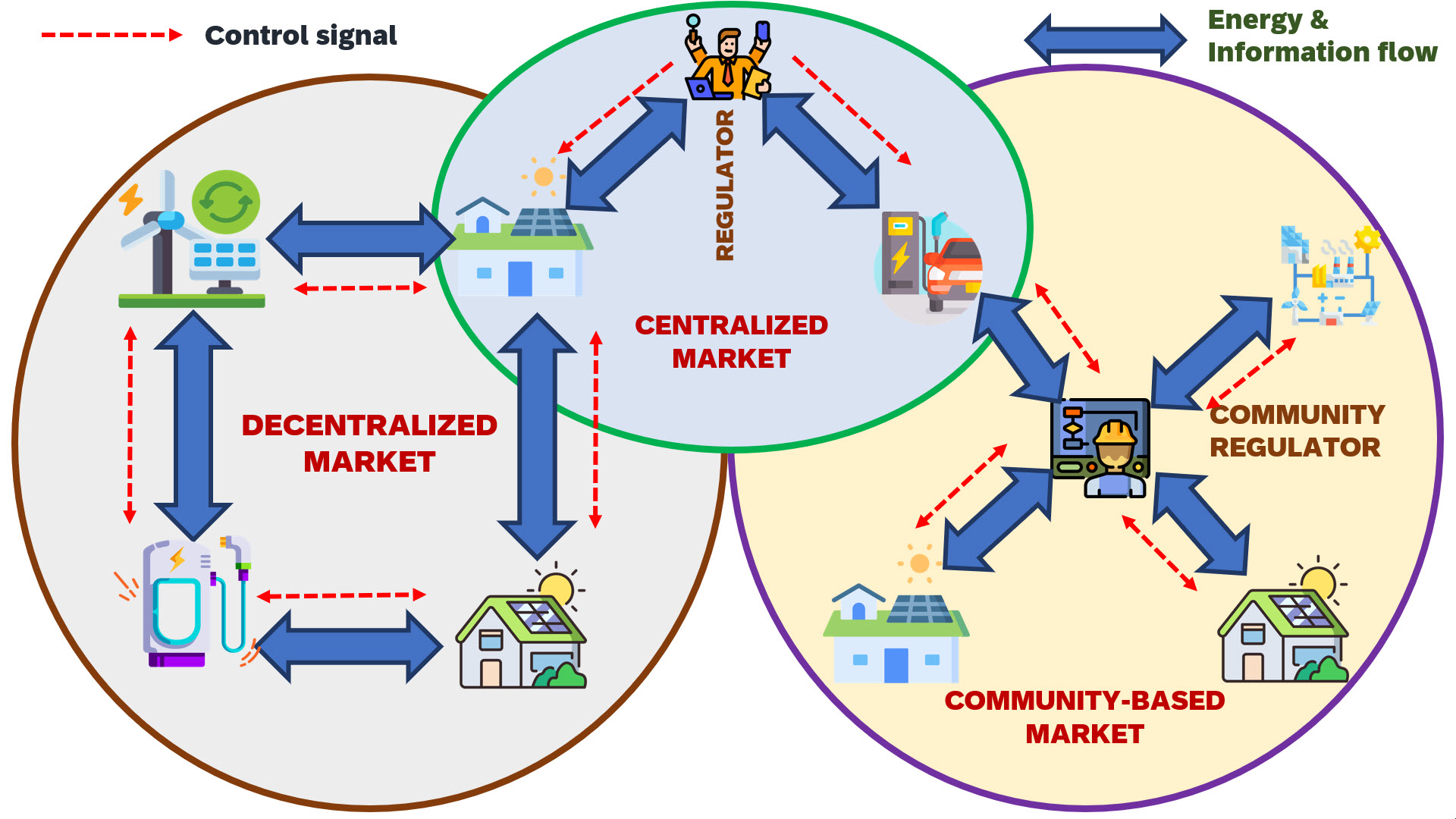}
    \caption{Market structures: Centralized market, Decentralized market and Community-based market}
    \label{fig:marketstructure}
\end{figure*}
\subsubsection{Pricing Mechanism}
Governments, energy providers, and utilities may use different energy-sharing pricing models. Energy sharing involves the exchange of energy resources between companies, with price policies designed to provide energy generators fair compensation and consumers affordable options. Pricing policies such as \ac{ToU} tariff, \ac{FiTs}, net metering, and Dynamic pricing are implemented by countries to solve the problems of consumers and prosumers related to \ac{DR}, load shifting. The following pricing policies are explained that are linked to energy sharing:

\begin{enumerate} [label=\roman*., itemsep=0pt, topsep=0pt]
    \item {\textbf{\ac{FiTs}}}:  \ac{FiTs} are pricing mechanisms that assign a fixed rate to the electricity that energy producers, particularly those producing renewable energy, input into the grid. Frequently established at a premium to the prevailing market rate, this rate encourages the implementation of renewable energy technologies. Energy sharing occurs through \ac{FiTs} when excess renewable energy is introduced into the power grid. It may also change how we support and utilize energy sources, with wider implications and effects on a variety of industries \cite{index}. The \ac{FiTs} program allows prosumers to buy grid power during peak demand and sell it back during low demand. However, their modest surplus electricity compensation is a downside. Some governments have ended or reconsidered certain projects due to financial constraints\cite{r1,r6}.
    \item {\textbf{Net metering}}: Net metering is a mechanism by which consumers who have installed on-site renewable energy systems, such as solar panels, are granted credits in return for the excess electricity they return to the infrastructure. When pricing is established on a retail rate basis, credits are allocated at the same rate at which an end user would remit payment for electricity delivered via the grid.
    \item {\textbf{\ac{ToU}}}: \ac{ToU} pricing structures are designed to account for fluctuating demand patterns by adjusting the cost of electricity. Amid periods of diminished demand, prices are reduced during off-peak hours. This incentivizes consumers to adjust their energy consumption to periods of reduced electricity costs. Although unrelated to energy sharing, \ac{ToU} pricing impacts the timing and manner in which consumers consume energy, thereby influencing the grid's overall dynamics.
    \item {\textbf{Real-time pricing}}:  Real-time pricing (RTP) of electricity is a dynamic pricing strategy in which the cost of electricity fluctuates in real-time and is determined by actual market conditions. Real-time pricing is in contrast to traditional fixed-rate pricing, which maintains a fixed value throughout a specified period. It is indicative of the current supply and demand in the electricity market. 
    \item {\textbf{Critical peak pricing}}: Critical Peak Pricing (CPP) is a pricing strategy for electricity whereby rates are increased for consumers during critical peak periods, which are occasions of heightened demand. CPP avoids the need for extra infrastructure to manage peak demand and reduces the burden on the electricity grid by encouraging consumers to reduce the use of electricity during these periods of increased demand. To align utility costs with the actual system conditions, CPP is a type of time-based pricing. 
    \item {\textbf{Wholesale electricity markets}}: Wholesale electricity markets facilitate the transaction between retailers and producers of electricity. The prices in these markets are dictated by the interplay between supply and demand. At market rates, producers distribute electricity to retailers, who then distribute it to end-users. Energy sharing takes place between various market entities at the wholesale level.
    \item {\textbf{\ac{PPAs}}}: \ac{PPAs} are agreements between an energy provider and a purchaser, which is typically a corporation or utility. The pricing structure of a \ac{PPAs} may consist of fixed, escalating, or market-rate-linked prices, among others. Commonly used for large-scale renewable energy initiatives, \ac{PPAs} may also require the grid to receive excess energy for sale.
    \item {\textbf{Capacity payments}}: In the absence of active electricity generation, generators are compensated for their capacity to generate energy through capacity payments. Maintaining grid reliability is a typical function of this mechanism. Capacity payments might constitute an element of the remuneration framework to facilitate energy sharing.
    \item {\textbf{Models of community-based pricing}}: Pricing models for community energy projects, such as community solar initiatives, might incorporate a subscription-based or collective ownership structure. The costs and benefits of the initiative are shared by all participants, and the pricing structure is intended to promote a fair and balanced distribution of the advantages.
    \item {\textbf{Blockchain technology and smart contracts}}: Among other emergent technologies, blockchain, and smart contracts offer decentralized and transparent mechanisms for conducting energy transactions. These technologies may facilitate peer-to-peer energy trading through the implementation of transparent, secure, and automated pricing mechanisms.

\end{enumerate}
\begin{table*}[]
\setlength{\tabcolsep}{10pt}
   \renewcommand{\arraystretch}{1.5}
    \setlength{\arrayrulewidth}{0.3mm}
    \centering
    \caption{Pricing Mechanisms}
    \label{tab:Pricingmechanism} 
    \begin{tabular}{|c|c|c|}
    \hline
    \textbf{Sl.No} & \textbf{Pricing mechanism}  & \textbf{References}\\
    \hline
    1 & \ac{FiTs}  &\cite{pyrgou2016future},\cite{couture2010policymaker},\cite{prahastono2019review},\cite{poruschi2018revisiting},\cite{alizamir2016efficient},\cite{8400476}\\
    \hline
    2 & Net metering  &\cite{poullikkas2013review},\cite{dufo2015comparative},\cite{christoforidis2016model},\cite{vieira2016net},\cite{darghouth2011impact},\cite{comello2017cost},\cite{eid2014economic},\cite{stoutenborough2008encouraging},\cite{yamamoto2012pricing},\cite{8400476}\\
    \hline
    3 & \ac{ToU} & \cite{carmichael2021demand},\cite{belton2020smart},\cite{sulaima2019review},\cite{li2019impact},\cite{nicolson2018consumer} \\
    \hline
    4 & Real Time Pricing & \cite{steriotis2018novel},\cite{tsaousoglou2019personalized},\cite{rahbari2016distributed},\cite{9252968},\cite{9495175},\cite{8836554},\cite{8467326},\cite{7893757},\cite{9663172} \\
    \hline
    5 & Critical Peak Pricing & \cite{9443080},\cite{6621016},\cite{7024930} \\
    \hline
    6 & Wholesale Electricity markets & \cite{8675533},\cite{9865182},\cite{8302928},\cite{9492792},\cite{8733097},\cite{9851640},\cite{9215970},\cite{9721654},\cite{9966489} \\
    \hline
     7 & Power Purchase Agreements & \cite{ghiassi2021making},\cite{ninomiya2020peer},\cite{7216645},\cite{9921036}\\
    \hline
    8 & Capacity Payments & \cite{4504976},\cite{5409538},\cite{soder2020review},\cite{bublitz2019survey}\\
    \hline
      \end{tabular} 
    \end{table*}
    
The various price mechanisms implemented in the electricity market, with a specific focus on energy sharing, are detailed in Table \ref{tab:Pricingmechanism}.  Finally, the structure of markets, regulatory frameworks, and specific goals of an Energy Sharing Initiative or Programme are often shaped by price policies in this sector. Their objective is to achieve equilibrium between energy producers' and consumers' concerns, as well as stability and effectiveness of the overall energy system.

\begin{table*}[]
\setlength{\tabcolsep}{10pt}
   \renewcommand{\arraystretch}{1.5}
    \setlength{\arrayrulewidth}{0.3mm}
    \centering
    \caption{Energy sharing: Case studies}
    \label{tab:Energy_sharing_case} 
    \begin{tabular}{|p{0.3cm}|p{1.9cm}|p{5.5cm}|p{2.5cm}|p{1.2cm}|p{0.3cm}|p{1cm}|}
    \hline
    \textbf{Sl.No} & \textbf{Project name} & \textbf{Details} & \textbf{Developed By} &  \textbf{Country} &  \textbf{Year} & \textbf{Reference}\\
    \hline
         1 & The Broooklyn Microgrid 
         &  \vspace{-1em} \begin{itemize}[leftmargin=*, nosep, topsep=0pt, partopsep=0pt]
            \item Community-driven network of New York City residents and business owners with locally generated solar energy. \end{itemize} \vspace{-1em} & LO3 energy company & USA & 2016 & \cite{mengelkamp2018designing}\\
        2 & Bornholm Island 
        &  \vspace{-1em} \begin{itemize}[leftmargin=*, nosep, topsep=0pt, partopsep=0pt]
            \item Project designed to fully harvest the offshore wind potential in the Baltic Sea to give power to Germany and  Denmark. 
            \item 3 GW wind energy is harvested, then converted to 525 kV HVDC and transported. \end{itemize} \vspace{-1em} & 50 Hertz and Energinet &  Germany \& Denmark &  2021 &  \cite{bornhlo} \\
        3 & sonnenCommunity & 
         \vspace{-1em} \begin{itemize}[leftmargin=*, nosep, topsep=0pt, partopsep=0pt]
            \item  Community members are generating their own electricity, storing it and sharing surpluses with friends or each other on the Internet. \end{itemize} \vspace{-1em} & sonnenBatterie & Germany & 2015 & \cite{Sonnen}  \\
        4 & Vandebron  
        &  \vspace{-1em} \begin{itemize}[leftmargin=*, nosep, topsep=0pt, partopsep=0pt]
            \item Align with the preferences of renewable energy providers and establish local clean energy communities. \end{itemize} \vspace{-1em}  & Vandebro & Netherlands &  2013 &  \cite{vandebron} \\
        5 & Piclo Flex  
        &  \vspace{-1em} \begin{itemize}[leftmargin=*, nosep, topsep=0pt, partopsep=0pt]
            \item A UK-based software platform for trading smart grid flexibility and P2P energy trading services. \end{itemize} \vspace{-1em} & UK Power Network & UK & 2013 &  \cite{Piclo} \\
        6 & AGL Virtual Trail  
        &  \vspace{-1em} \begin{itemize}[leftmargin=*, nosep, topsep=0pt, partopsep=0pt]
            \item Identify and value P2P energy trades, determine the applicability of distributed ledger technology, and analyze the results. \end{itemize} \vspace{-1em} & AGL Energy Services Pty Ltd & Australia & 2017 &  \cite{AGL} \\
        7 & KEPCO  
        &  \vspace{-1em} \begin{itemize}[leftmargin=*, nosep, topsep=0pt, partopsep=0pt]
            \item  By using a blockchain system that is enabled, solar energy suppliers are able to deliver more electricity to their customers. \end{itemize} \vspace{-1em}  & Kansai Electric Power Company (KEPCO) and Power Ledger & Japan & 2018  &  \cite{kepco} \\
        8 & Pilot project in Delhi  
        &  \vspace{-1em} \begin{itemize}[leftmargin=*, nosep, topsep=0pt, partopsep=0pt]
            \item Power Ledger's blockchain technology facilitates P2P solar energy trading from over 2 MW of solar photovoltaic systems.
            \item The pilot is currently underway among 65 consumers and 75 consumption locations in the capital - 140 buildings and counting in New Delhi. \end{itemize} \vspace{-1em} & Tata Power-DDL & India & 2021 &  \cite{India} \\
        9 & Pilot project in Uttar Pradesh 
        &  \vspace{-1em} \begin{itemize}[leftmargin=*, nosep, topsep=0pt, partopsep=0pt]
            \item  The government of Uttar Pradesh, which is the only Indian state to amend its regulatory framework allowing regulated peer to peer energy trading, has introduced blockchain technology in rooftop solar. \end{itemize} \vspace{-1em}  & Uttar Pradesh Power Corporation (UPPCL) and Uttar Pradesh New and Renewable Energy Development Agency (UPNEDA) & India & 2019 &  \cite{India2} \\
        10 & A Korean project 
        &  \vspace{-1em} \begin{itemize}[leftmargin=*, nosep, topsep=0pt, partopsep=0pt]
            \item  In order to demonstrate the benefits of flexibility trading on Korea's market when decarbonisation is underway, a United Kingdom blockchain startup will operate an Energy Flexible Trading Platform in South Korea. \end{itemize} \vspace{-1em}  & Electron & Korea & 2018 & \cite{korea} \\
        \hline 
      \end{tabular} 
    \end{table*}

\begin{table*}
\setlength\extrarowheight{2mm}
    \centering
    \caption{Double Auction Theory with Blockchain Technology.}
  \label{tab:Double auction & Blockchain}
    \begin{tabular}{|p{2cm}|p{3.5cm}|p{9cm}|p{1.4cm}|}
    \hline
  \textbf{Type} & \textbf{Methodology } & \textbf{Contribution} & \textbf{Reference} \\
  \hline
      Continuous 
      &  Iterative 
      & \vspace{-1em} \begin{itemize}[leftmargin=*, nosep, topsep=0pt, partopsep=0pt]
            \item  Energy allocation and market equilibrium shall be ensured in an optimum way.
             \item The autonomous actualization of the developed \ac{P2P} energy trading model on the blockchain platform.
        \end{itemize} \vspace{-1em}
& \cite{zhang2021peer}\\
    &  $k$-factor  
    & \vspace{-1em} \begin{itemize}[leftmargin=*, nosep, topsep=0pt, partopsep=0pt]
            \item Benefits of the buyers and seller feasibility were studied by varying $k$ value.
          \end{itemize} \vspace{-1em}
    & \cite{angaphiwatchawal2020k}\\
    &  Prediction integration strategy optimization (PISO) model  
    & \vspace{-1em} \begin{itemize}[leftmargin=*, nosep, topsep=0pt, partopsep=0pt]
            \item Prosumers operations and market trading strategies are optimized.
    \end{itemize} & \cite{chen2019trading}\\
&  Multi-$k$   & \vspace{-1em} \begin{itemize}[leftmargin=*, nosep, topsep=0pt, partopsep=0pt]
    \item It allows participants to negotiate exchange prices.
\end{itemize} \vspace{-1em}  & \cite{angaphiwatchawal2021multi}\\
& Ethereum and uniform price mechanism 
& \vspace{-1em} \begin{itemize}[leftmargin=*, nosep, topsep=0pt, partopsep=0pt]
            \item The design was validated and compared by conducting A/B tests. This complements the current centralized energy grid.  
            \end{itemize} \vspace{-1em} & \cite{vieira2021peer}\\
& Q-learning  
& \vspace{-1em} \begin{itemize}[leftmargin=*, nosep, topsep=0pt, partopsep=0pt]
            \item  The study was carried out in Guizhou Province of China for 14 microgrids.
            \item  The proposed Qcube methodology allows rational decisions to be made, as well as maximizing the profits of microgrids.  
            \end{itemize} \vspace{-1em} & \cite{wang2019q}\\
& Consortium blockchain 
& \vspace{-1em} \begin{itemize}[leftmargin=*, nosep, topsep=0pt, partopsep=0pt]
            \item The proposed method reduces transaction costs and improves efficiency. 
            \item  In a continuous double auction, fair blind signing technology is generating pseudonyms and certificate of anonymity that allow identity to be kept private while maintaining decentralization.
            \end{itemize} \vspace{-1em} & \cite{zhang2019privacy}\\
\hline
Periodic & Local energy market model 
& \vspace{-1em} \begin{itemize}[leftmargin=*, nosep, topsep=0pt, partopsep=0pt]
            \item  The algorithm assures user preferences, considers willingness to pay, increases local coverage of electricity, maintains sanity, and is computationally tractable. 
            \end{itemize} \vspace{-1em} & \cite{zade2022satisfying} \\
& Multi-agent system (MAS) architecture 
& \vspace{-1em} \begin{itemize}[leftmargin=*, nosep, topsep=0pt, partopsep=0pt]
            \item  Integrated 5G and edge computing with a security-aware environment analyzed with STRIDE threat modeling.  
            \end{itemize} \vspace{-1em} & \cite{kalbantner2021p2pedge} \\
& Community energy market model   
& \vspace{-1em} \begin{itemize}[leftmargin=*, nosep, topsep=0pt, partopsep=0pt]
            \item  A balanced approach has been adopted to address the Community's budget deficit due to time differences between supply and demand. 
            \item Uncertainties were addressed and the proposed model was validated using real-market household data.  \end{itemize} \vspace{-1em}  & \cite{alabdullatif2020market} \\
\hline
Smart contracts & Ethereum-based 
& \vspace{-1em} \begin{itemize}[leftmargin=*, nosep, topsep=0pt, partopsep=0pt]
            \item  Smart contracts were written in Solidity and used Remix IDE to deploy them. Data on an energy market scenario have been selected to test these contracts.  
            \end{itemize} \vspace{-1em} & \cite{damisa2022towards} \\
&  Consortium blockchain  
& \vspace{-1em} \begin{itemize}[leftmargin=*, nosep, topsep=0pt, partopsep=0pt]
            \item  In real time, a physical demonstration project in Seattle is simulated for the simulation of power grid scenarios. 
            \end{itemize} \vspace{-1em} & \cite{foti2019blockchain} \\
\hline
Dynamic pricing & Ethereum-based 
& \vspace{-1em} \begin{itemize}[leftmargin=*, nosep, topsep=0pt, partopsep=0pt]
            \item  A smart contract resides on the blockchain to guarantee that trade is executed in a precise fashion, maintains an immutable transaction record, and eliminates high costs and overhead.
            \item  Moreover, it prevents the double sale and creates a dynamic pricing model.  
            \end{itemize} \vspace{-1em} & \cite{song2021smart}\\ 
\hline
Decentralized & Flocking-based &  \vspace{-1em} 
        \begin{itemize}[leftmargin=*, nosep, topsep=0pt, partopsep=0pt]      
        \item Guaranteed \ac{P2P} trading between prosumers within the neighborhood. The method preserves the anonymity of the buyer and seller.
        \end{itemize}
        \vspace{-1em}
        &   \cite{bandara2021flocking}\\
& Distributed  & 
 \vspace{-1em} 
        \begin{itemize}[leftmargin=*, nosep, topsep=0pt, partopsep=0pt]
        
        \item Maximization of the preferences and needs of the peers by optimal prosumer scheduling while combining demand response with the decentralized network.
        \end{itemize}
        \vspace{-1em}
        & \cite{kalakova2021blockchain}\\
& ETradeChain  &        
 \vspace{-1em} 
        \begin{itemize}[leftmargin=*, nosep, topsep=0pt, partopsep=0pt]
        \item The ETradeChain utilized a double auction process to reach a consensus on energy transactions while employing blockchain technology to demonstrate real-time P2P trading.
        \end{itemize} \vspace{-1em} & \cite{barbhaya2023etradechain} \\
\hline
\end{tabular}
\end{table*}

\subsection{Energy Sharing: Case Studies}
Energy-sharing projects have been launched across various parts of the globe to demonstrate \ac{P2P} energy sharing. In particular, countries in North America, Europe, Australia, and Asia are heavily involved in studies in various testing facilities, as discussed in \cite{tushar2021peer}. A comparison of the major \ac{P2P} electricity trading cases was analyzed in \cite{PARK20173}.  A few examples of energy-sharing projects that have been implemented in various parts of the world are given in Table \ref{tab:Energy_sharing_case}.

\section{Benefits of Energy sharing}
Energy sharing transforms sustainability and resilience by providing many benefits. A major benefit is resource optimization. Efficiency and waste reduction are achieved by spreading excess energy from surplus regions to deficit regions. This collaborative concept ensures energy system dependability and cuts costs. Energy market players use surplus energy during peak demand periods to save money on redundant infrastructure. Energy sharing promotes environmental sustainability by facilitating the integration of renewable energy into the grid. Renewable-rich regions share clean energy with more impoverished ones, promoting a greener power transition. Flexible energy sharing accommodates dynamic demand and renewable resource availability, boosting adaptation. Innovation in grid management and smart energy systems is spurred by energy sharing. Sharing energy improves energy security, reduces transmission losses, and promotes global cooperation, creating a resilient, sustainable, and linked energy future.

\subsection{Enhancement of Energy Security}
A varied and integrated energy management approach through energy sharing improves energy security. One major benefit is energy independence. Regions with ample energy supplies help those with shortages through energy-sharing methods, reducing the risks of over reliance on specific fuels or regions. Sharing energy increases resilience to disturbances. Sharing resources makes an energy system resilient to catastrophes caused by nature, international conflicts, and other unforeseen circumstances. During uncertain circumstances, excess energy from one location may compensate for shortcomings in another, preserving the power supply. 

Energy security is improved via energy sharing and distributed energy resource integration. Localized generation like solar panels, wind turbines, and microgrids reduces large-scale failures. Distributed resources power key infrastructure and essential services during grid failures, making energy infrastructure more resilient and secure. To evaluate the structure of the electricity market and the spot price as traditional energy sources (such as fossil fuels and hydropower) decline and renewable energy sources (such as solar and wind) rise, a system dynamics model is presented in \cite{rios2021renewable}. In a day-ahead market, energy metrics including reserve margin, resilience, dependability, and vulnerability are used to quantify energy security. Energy sharing promotes strategic regional and international partnerships. International energy pooling agreements will boost geopolitical stability by encouraging mutual advantages. Energy conflicts will be avoided by working together to provide a constant energy supply in countries with various energy resources. For energy security, energy sharing diversifies sources, increases resilience to disturbances, promotes distributed energy supplies, and encourages international cooperation. Energy sharing stabilizes the global energy landscape by connecting and adapting energy networks.

\subsection{Energy Resilience Improvement}
Energy resilience is a system's ability to resist, adapt to, and quickly recover from disruptions while maintaining a steady energy supply. This notion is vital for assuring energy resource availability and functionality in the face of natural catastrophes, cyberattacks, equipment failures, and other unforeseen events. Energy sharing boosts energy ecosystem resilience. Enhancing resilience with energy sharing has various benefits.
Shuhan Yao et. al \cite{8334269} proposed a time-space Transportable Energy Storage (TESS) model covering both the transport network and \ac{EDS}  to demonstrate the difference between TESS and \ac{ESS} in terms of flexibility and cost reduction of sharing \ac{ESS}  between microgrids. The proposed recovery problem is formulated as a mixed integer linear programming considering different networks and their TESS constraints. The proposed model and scheme are tested on a modified 33-bus test system with three microgrids and four TESSs.The results confirmed that distributed systems with TESS are more resilient than traditional \ac{ESS}, as benefits from overall cost savings. The resilience to communication errors in reactive power-sharing control of an AC MG based on an inverter is depicted in \cite{9359673}. In the event of communication failures, precise reactive power sharing is ensured by creating an adaptive resilient control scheme.
To enable networked microgrids to organize effectively for extreme occurrences and fully adjust to subsequently changing conditions, which improves the system's resilience, flexible division, and unification control mechanisms are proposed in \cite{8786210}. The suggested techniques indicate that networked microgrids transition between two divisions and unification modes through a sparse communication network without needing extra controllers or communication infrastructures. The planned deployment of mobile emergency resources (MERs) such as s hydrogen refueling stations (HRSs) and mobile fuel cell vehicles (MFCVs) enhances the capacity of \ac{EDS} to self-heal in the case of extreme catastrophes \cite{9770398}. A  P2P energy exchange system is presented in \cite{spiliopoulos2022peer} to help microgrids operate more economically and robustly, the proposed not only has economic benefits and reduces carbon emissions but also improves resilience by up to 80 percent. 

To address the unfavorable conditions that an \ac{EDS} with integrated DER may encounter, Divyanshi Dwivedi et al.\cite{dwivedi2023evaluating} suggested a hybrid data-driven approach that utilizes machine learning and complex networks. This technique examined the system resilience by rendering the real power among electrical loads and linking it with correlated networks for analysis. In addition, the reliability and resilience of the network were increased by installing \ac{DER}s on the weaker nodes. By calculating the percolation threshold for the microgrid networks, resilience improvements are verified and the results showed that there is an improvement of 20.45\% as a result of the optimal placement of \ac{DER}s \cite{reddy2022data}. The study conducted by Divyanshi Dwivedi et al.\cite{dwivedi2023advancements} examined the methods used recently to improve resilience, including operational and planning strategies. A thorough description of \ac{P2P} energy trading, renewable energy integration, and microgrid deployment in strengthening power networks against disturbances is given. 

\subsection{Social Welfare Benefits}
It is a good practice for society to share energy to promote collaboration, justice, and sustainability in the energy sector. An atmosphere of shared responsibility and mutual support has been created within the communities by adopting the concept of energy sharing. As individuals, businesses, and organizations are sharing resources of energy, this cooperation approach is having a positive impact on Social Cohesion. Sharing energy also helps communities and microgrids reduce their carbon emissions by promoting the use of renewable energy sources. Besides environmental benefits, the positive effects of energy sharing extend to customers because they are provided with additional electricity thanks to \ac{P2P} trading and locally regulated energy markets while offering opportunities for prosumers to generate income by selling surplus power. The benefits of sustainable practices become accessible to a larger portion of the population owing to this diversification of energy generation and distribution. Furthermore, energy sharing acts as a catalyst for upgrading the standard of life as well as energy availability in regions with limited access to centralized power systems. By utilizing the alternate direction method of multipliers (ADMM) algorithm, Shaobo Yang et al.\cite{9931111} suggested energy sharing mechanism for ADN to increase the local accommodation degree of renewable energy from 84.85\% to 85.27\% while maximizing social welfare. and it rises in tandem with the number of prosumers who install energy storage in their homes. Resilient, inclusive, and ecologically conscious societies are fostered when communities actively engage in and realize the benefits of energy-sharing schemes.

\section{\ac{P2P} Energy trading-Market Clearing Algorithms}
The broader power grid encompassing generators, the distribution system, and retailers stand to gain substantial advantages from this innovative approach. P2P trading holds the potential to reduce reserve requirements, lower investment and operational costs, mitigate peak demand, and enhance grid reliability. However, despite these benefits, P2P electricity trading encounters several challenges. Firstly, the absence of a central coordinator introduces reliability concerns, necessitating the establishment of trust among prosumers. Furthermore, modeling the decision-making processes involving numerous participants with conflicting interests is inherently complex. Fully decentralized P2P trading systems may also pose security risks to the network. Lastly, grid stakeholders may have varied demands for P2P services from prosumers with differing objectives, requiring innovative pricing mechanisms to prevent network congestion.

Numerous approaches to addressing these problems have been explored in the literature. Gaining a comprehensive understanding of the entire \ac{P2P} electricity market may be challenging due to the inherent complexity and diversity of these systems. However, understanding recent work in this area is crucial for new lines of inquiry, addressing new issues in the energy sector, and creating more \ac{P2P} electricity trading services. Several scholars have contributed their distinct viewpoints on \ac{P2P} trading through their insightful reviews, providing insightful analysis and knowledge in this rapidly evolving field \cite{soto2021peer},\cite{bukar2023peer},\cite{tushar2020peer},\cite{sachan2022charging},\cite{al2021charging},\cite{das2023peer},\cite{ullah2023electric}. Nevertheless, the breadth of these reviews is constrained because they don't fully address ongoing concepts in the literature.

\begin{figure*}
    \centering
    \includegraphics[width=1\linewidth]{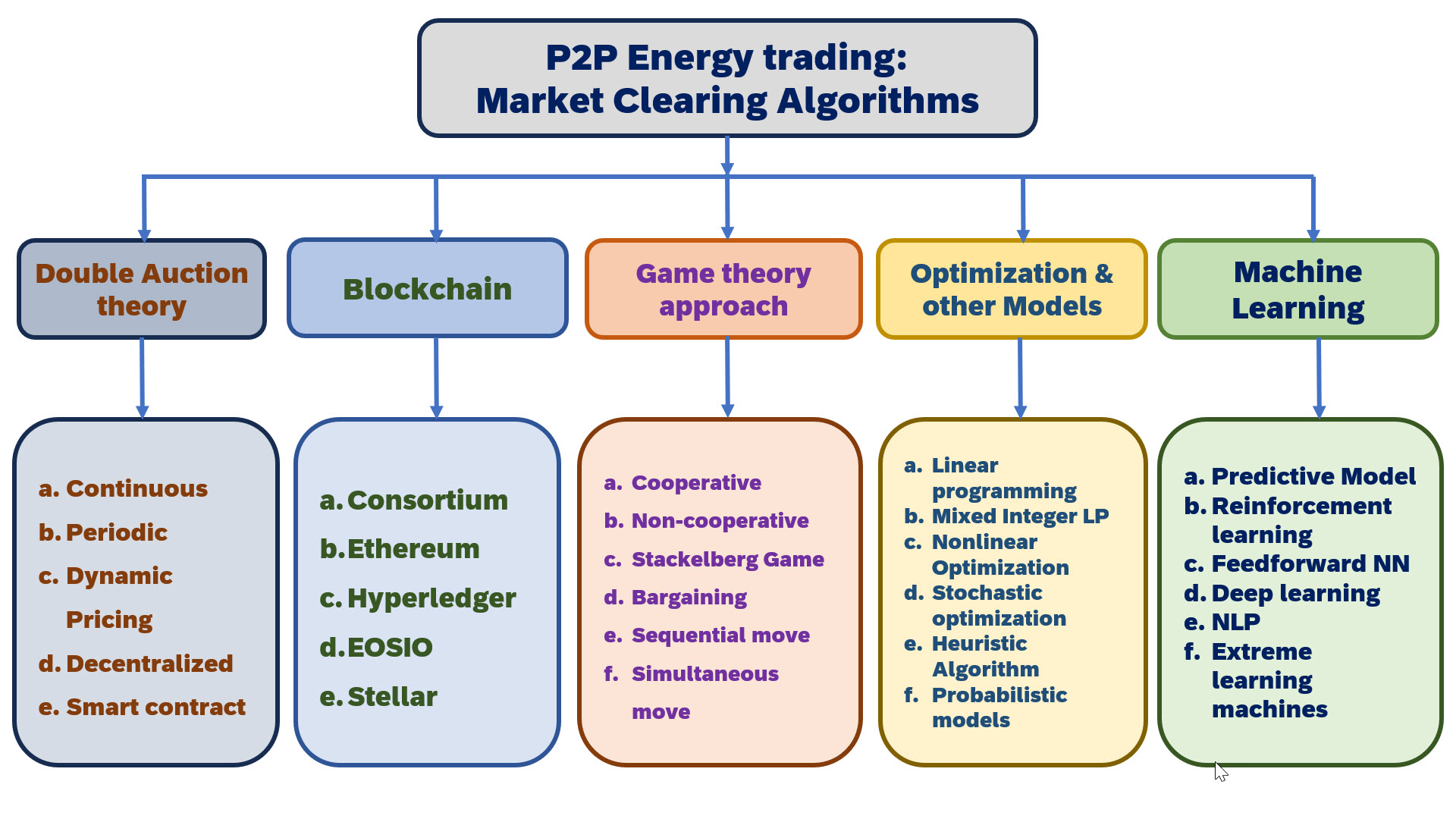}
    \caption{P2P Energy Sharing-Market Clearing Algorithms}
    \label{fig:ETMCA}
\end{figure*}

\ac{P2P} energy trading facilitates the interaction between different participants in the electricity market. It will allow for direct energy purchases and sales between individuals and organizations in a way that does not involve intermediaries. This helps to create a more efficient and cost-effective energy market while promoting renewable energy adoption and reducing carbon footprint. The mathematical model for fair transactions is designed based on players' behavior. The literature reports five formulation approaches: double auction theory, blockchain technology, game theory, models \& optimization algorithms, and machine learning, as given in Fig. \ref{fig:ETMCA}.
\subsection{Double Auction Theory with Blockchain Technology}
\begin{figure}
    \centering
    \includegraphics[width=1\linewidth]{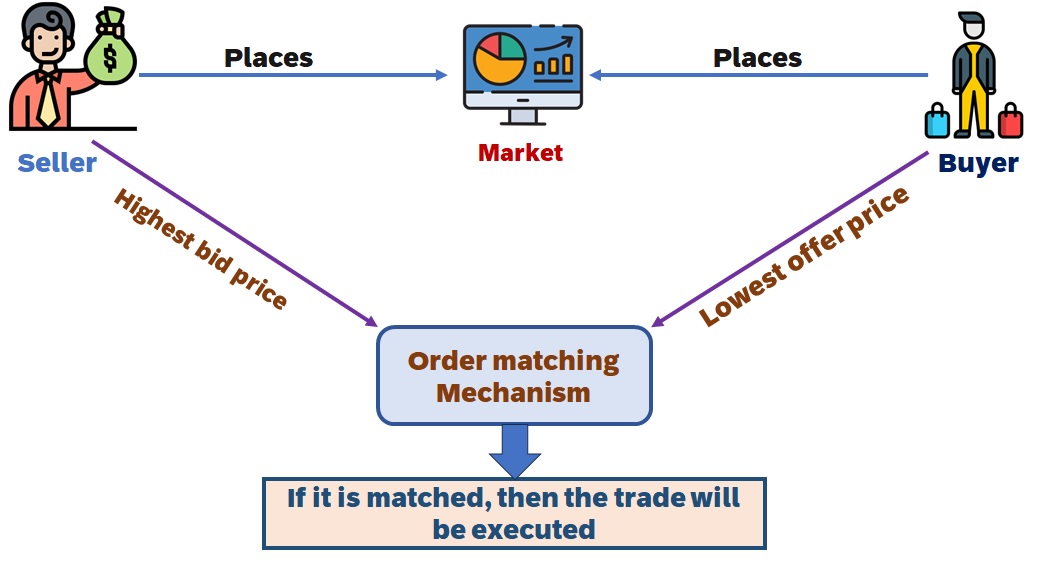}
   \caption{Double Auction Theory}
    \label{fig:DAT}
\end{figure}

Double auction theory is a kind of trading mechanism in which buyers and sellers in an auction market submit competitive bids and offers simultaneously. Fig \ref{fig:DAT} illustrates the process of double action theory in the electricity market where buyers and sellers participate in the auction. The price at which a stock trades indicates the maximum amount that a buyer is prepared to spend and the minimum amount that a seller is prepared to take. Orders are performed after matching bids and offers are linked together, and trading ceases when the match is not found. Double Auction theory is classified into Continuous, Periodic, Multilateral, Smart Contract, Dynamic pricing, Decentralized, Retail and Wholesale, and Localized. Literature on Double auction theory is given in \cite{friedman2018double}, meanwhile \cite{klemperer1999auction} addresses the complete analysis of the auction market with some examples. The double auction theory is used along with blockchain technology for energy trading between peers in the electricity market. Literature found promising articles for energy trading and is presented and illustrated in Table \ref{tab:Double auction & Blockchain}.

\begin{table*}
\setlength\extrarowheight{1mm}
\renewcommand{\arraystretch}{1}
    \centering
     \caption{Blockchain Technology- Consortium}
  \label{tab:Blockchain-consortium}
    \begin{tabular}{|p{1.2cm}|p{14cm}|}
    \hline
   \textbf{Reference} & \textbf{Contribution} \\
  \hline
\cite{li2017consortium} &\vspace{-1em} \begin{itemize}[leftmargin=*, nosep, topsep=0pt, partopsep=0pt] 
 \item  A credit based payment scheme and a best pricing strategy using the Stackelberg game to handle credits load have been suggested. \end{itemize} \vspace{-1em} \\
\cite{aggarwal2021consortium} & \vspace{-1em} \begin{itemize}[leftmargin=*, nosep, topsep=0pt, partopsep=0pt] \item  A double auction theory for maximizing social welfare has solved the energy pricing problem and the level of traded energy problems that arise in response to demand. \end{itemize} \vspace{-1em} \\
 \cite{yang2020compensation} & \vspace{-1em} \begin{itemize}[leftmargin=*, nosep, topsep=0pt, partopsep=0pt]
      \item A blockchain process and smart contracts were established.
      \item The creation of a type of cryptocurrency called ``elecoin" in a P2P market is realized using a case study. 
  \end{itemize} \vspace{-1em} \\
\cite{kang2017enabling} & \vspace{-1em} \begin{itemize}[leftmargin=*, nosep, topsep=0pt, partopsep=0pt]
      \item Detailed operation of localized P2P electricity trading is illustrated.
      \item  An iterative double auction mechanism to optimize social benefits solves the issue of electricity prices and offers trading between \ac{PHEV}.
  \end{itemize} \vspace{-1em}  \\
\cite{che2019distributed} & \vspace{-1em} \begin{itemize}[leftmargin=*, nosep, topsep=0pt, partopsep=0pt]
      \item A certification authority is introduced to achieve privilege control and supervision of transaction parties.
      \item The process of transaction authentication is simulated in consortium blockchain, and Hyperledger caliper is used for evaluation of the above model.
  \end{itemize} \vspace{-1em} \\
 \cite{khalid2021consortium} & \vspace{-1em} \begin{itemize}[leftmargin=*, nosep, topsep=0pt, partopsep=0pt]
     \item Smart contracts are designed to ensure fair payment, and an algorithm is developed for energy trading for \ac{EV}s.
     \item  The power flows and associated power losses are discussed.
\end{itemize} \vspace{-1em} \\
\cite{fu2020intelligent} & \vspace{-1em} \begin{itemize}[leftmargin=*, nosep, topsep=0pt, partopsep=0pt]
    \item Charging information is managed by consortium blockchian and To balance the company's profit a new smart contract is designed and to support this LNSM is proposed.
    \item  An application to a real \ac{EV} charging case study confirms the proposed framework and smart contract.
\end{itemize} \vspace{-1em} \\
\cite{abishu2021consensus}  &  \vspace{-1em} \begin{itemize}[leftmargin=*, nosep, topsep=0pt, partopsep=0pt]
     \item  In the case of electricity trading, validator selection, block generation and consensus operations shall be carried out in every cluster within a vehicular network..\item An incentive mechanism based on a Stackelberg game model is proposed to optimize the utility of sellers, buyers, and validator nodes, which motivates honest and cooperative nodes.
\end{itemize} \vspace{-1em}  \\
\cite{li2021electric} &  \vspace{-1em} \begin{itemize}[leftmargin=*, nosep, topsep=0pt, partopsep=0pt]
     \item A SMES is introduced to store surplus electric energy. and Distribution is adjusted intelligently, and blockchain is applied.
     \item \ac{P2P} electricity trading model is realized by the Cooperative game.
\end{itemize} \vspace{-1em}   \\
\cite{li2019iterative} & \vspace{-1em}  \begin{itemize}[leftmargin=*, nosep, topsep=0pt, partopsep=0pt]
     \item Initially, the overall load variance of the distribution network is minimized using a heuristic algorithm.
     \item The krill herd algorithm is proposed, and finally decentralized trading architecture is designed.
\end{itemize} \vspace{-1em} \\
\cite{9311120} &  \vspace{-1em} \begin{itemize}[leftmargin=*, nosep, topsep=0pt, partopsep=0pt]
      \item  A method for the generation of the account using a time series single exponential technique is put in place.
      \item  The security analysis of the Smart Contracts model is provided and its performance is measured in simulations.
\end{itemize} \vspace{-1em} \\ 
\cite{9091029} & \vspace{-1em}  \begin{itemize}[leftmargin=*, nosep, topsep=0pt, partopsep=0pt]
     \item Blockchain-enabled architecture is used to facilitate energy trading between \ac{EV}s and critical loads in a microgrid.\item To monitor the energy trading activities between entities remotely, an energy trading prototype has been developed.
\end{itemize} \vspace{-1em}  \\
\cite{8457186} & \vspace{-1em} \begin{itemize}[leftmargin=*, nosep, topsep=0pt, partopsep=0pt]
     \item Permission energy blockchain system is introduced, then a reputation-based delegated Byzantine fault tolerance consensus algorithm is proposed.\item  Optimisation of contracts is analyzed in the light of contract theory. In addition, a new mechanism for the allocation of energy is suggested.
\end{itemize} \vspace{-1em}   \\
\cite{8889684} & \vspace{-1em} \begin{itemize}[leftmargin=*, nosep, topsep=0pt, partopsep=0pt]
     \item A decentralized power trading model is designed.  \item The reverse auction mechanism-based dynamic pricing strategy is deployed.\item \ac{V2G} \ac{EV} power trading smart contract is implemented.
\end{itemize} \vspace{-1em}  \\
\hline 
\end{tabular}
\end{table*}

\begin{table*}
\setlength\extrarowheight{2mm}
\renewcommand{\arraystretch}{1}
    \centering
     \caption{Blockchain Technology-Ethereum and Hyperledger}
  \label{tab:Blockchain_ethereum}
    \begin{tabular}{|p{1.3cm}|p{13cm}|p{1.2cm}|}
    \hline
  \textbf{Method}  & \textbf{Contribution} & \textbf{Reference} \\
  \hline
Ethereum  
& \vspace{-1em} \begin{itemize}[leftmargin=*, nosep, topsep=0pt, partopsep=0pt]
            \item  The energy trading is executed in a decentralized manner by leveraging blockchain technology. 
            \item The auction models are proposed for energy trading by smart contracts.
\end{itemize} \vspace{-1em}  & \cite{8245449}\\
 
& \vspace{-1em} \begin{itemize}[leftmargin=*, nosep, topsep=0pt, partopsep=0pt]
            \item Designed a transactive energy framework for a community microgrid. \item Implemented binary genetic algorithm for optimal scheduling of the loads. \item A prototype of the proposed model is designed in the laboratory and tested.   \end{itemize} \vspace{-1em} & \cite{9535388}\\

& \vspace{-1em} \begin{itemize}[leftmargin=*, nosep, topsep=0pt, partopsep=0pt]
            \item For effective \ac{P2P} trading, a modular smart contract mechanism is proposed within \ac{VPP} framework, which was adapted and converted to decentralized applications. \end{itemize} \vspace{-1em} & \cite{9204678}\\
 
& \vspace{-1em} \begin{itemize}[leftmargin=*, nosep, topsep=0pt, partopsep=0pt]
            \item Designed two frameworks namely smart contract functionality in Ethereum, and continuous double auction and uniform-price double-sided auction. \end{itemize} \vspace{-1em} & \cite{vieira2021peer}\\
 
& \vspace{-1em} \begin{itemize}[leftmargin=*, nosep, topsep=0pt, partopsep=0pt]
            \item The Ethereum smart contracts are designed and evaluated considering economic and technical constraints. \end{itemize} \vspace{-1em}  & \cite{kumari2022blockchain}\\
& \vspace{-1em} \begin{itemize}[leftmargin=*, nosep, topsep=0pt, partopsep=0pt]
        \item The proposed trading platform contemporary billing scheme.\item To check the efficient energy allocation, the proposed billing scheme is compared with existing schemes.\item Validation of the proposed trading platform that supports growing number of \ac{EV}s is done on the framework based on lightweight virtualization.
\end{itemize} \vspace{-1em}  & \cite{9125997}\\

& \vspace{-1em} \begin{itemize}[leftmargin=*, nosep, topsep=0pt, partopsep=0pt]
            \item  An AC-OPF problem is converted to a single optimization problem.\item Ethereum blockchain technology to fulfill the role of a virtual aggregator.\item  Using the Amsterdam prosumer community's database, the model was tested.
\end{itemize} \vspace{-1em} & \cite{van2020integrated}\\
\hline
Hyperledger .  
& \vspace{-1em} \begin{itemize}[leftmargin=*, nosep, topsep=0pt, partopsep=0pt]
            \item To establish trust between users, it is using Blockchain technology. Smart contracts were used for the payments electronically. \end{itemize} \vspace{-1em} & \cite{khan2021blockchain}\\
& \vspace{-1em} \begin{itemize}[leftmargin=*, nosep, topsep=0pt, partopsep=0pt] 
\item Two noncooperative games with dynamic supplier pricing, the suggested demand–response mechanism is established. \item An energy trading system is prototyped on a cluster network with a Hyperledger blockchain smart contract coordinator. \end{itemize} \vspace{-1em} & \cite{9542944}\\
& \vspace{-1em} \begin{itemize}[leftmargin=*, nosep, topsep=0pt, partopsep=0pt] \item  Docker and Go simulate a credit-based P2P electricity trading model on Hyperledger Fabric.  \item The experiments have shown that the suggested model on the Hyperledger blockchain reduces user costs and improves trading stability and efficiency in P2P electricity trading while managing credit. \end{itemize} \vspace{-1em} & \cite{9531949}\\

\hline
\end{tabular}
\end{table*}

\begin{table*}
\setlength\extrarowheight{2mm}
\renewcommand{\arraystretch}{1}
    \centering
     \caption{Game Theory Approaches}
  \label{tab:GT}
    \begin{tabular}{|p{1.3cm}|p{13cm}|p{1.2cm}|}
    \hline
  \textbf{Type} & \textbf{Contribution} & \textbf{Reference} \\
  \hline
  Cooperative game theory  & \vspace{-1em} \begin{itemize}[leftmargin=*, nosep, topsep=0pt, partopsep=0pt] \item The time-varying production of the hybrid wind power and PV power is implemented by multi-objective optimization and  \ac{P2P} trading is carried out for better results among nanogrid.  \end{itemize} \vspace{-1em}  & \cite{9799005}\\
  & \vspace{-1em} \begin{itemize}[leftmargin=*, nosep, topsep=0pt, partopsep=0pt] \item Myerson value rule distributes the proposed game's payment fairly among prosumers. \item The simulation results show that the proposed model avoids high voltage problems and also reduces electricity prices.   \end{itemize} \vspace{-1em}  & \cite{9392022}\\
  & \vspace{-1em} \begin{itemize}[leftmargin=*, nosep, topsep=0pt, partopsep=0pt] \item Myerson value rule distributes the proposed game's payment fairly among prosumers. \item The simulation results show that the proposed model avoids high voltage problems and also reduces electricity prices.   \end{itemize} \vspace{-1em}  & \cite{9392022}\\
& \vspace{-1em} \begin{itemize}[leftmargin=*, nosep, topsep=0pt, partopsep=0pt] \item The scheduling problem of microgrids is solved by the cooperative game theory approach in the presence of \ac{ESS}s, \ac{DR}s program and \ac{EV}s.  \end{itemize} \vspace{-1em}  & \cite{seyyedi2022stochastic}\\
& \vspace{-1em} \begin{itemize}[leftmargin=*, nosep, topsep=0pt, partopsep=0pt] \item Computed the percolation threshold with the help of an energy prediction model using ARIMA to show that there is improvement of energy resilience.  \end{itemize} \vspace{-1em}  & \cite{babu2023resilient}\\
& \vspace{-1em} \begin{itemize}[leftmargin=*, nosep, topsep=0pt, partopsep=0pt] \item a cooperative energy market model using the Generalized Nash Bargaining (GNB) theory for an active Distribution Network (DN). \item The proposed one greatly boosts social welfare through Volt-VAR control and maximizes profit allocation fairness under price restrictions.  \end{itemize} \vspace{-1em}  & \cite{9223743}\\
  \hline
  Non-cooperative game theory  
  & \vspace{-1em} \begin{itemize}[leftmargin=*, nosep, topsep=0pt, partopsep=0pt] \item The proposed trading scheme achieved a reduction in the energy bills of the consumers which is evaluated on 14 bus systems with 8 producers,  and 11 consumers.  \end{itemize} \vspace{-1em}  & \cite{amin2020motivational}\\
  & \vspace{-1em} \begin{itemize}[leftmargin=*, nosep, topsep=0pt, partopsep=0pt] \item Developed a novel technique with the least amount of information overhead for stable and equitable energy sharing among MG clusters. \item In addition to stability and fairness, the comprehensive numerical analysis validates the superiority of the suggested approach. \end{itemize} \vspace{-1em}  & \cite{8316917}\\
  & \vspace{-1em} \begin{itemize}[leftmargin=*, nosep, topsep=0pt, partopsep=0pt] \item Examined energy-efficient building management for a cluster of office, industrial, and commercial buildings with distributed transactions and the executed simulation results suggested energy sharing technique is computationally efficient, profitable.  \end{itemize} \vspace{-1em}  & \cite{8669964}\\
  & \vspace{-1em} \begin{itemize}[leftmargin=*, nosep, topsep=0pt, partopsep=0pt] \item Formulated stylized model for storage sharing, spot market with random clearing prices and investment decision are modeled. \item The results admit single Nash equilibrium and promote social welfare. \end{itemize} \vspace{-1em}  & \cite{8025410}\\
  \hline
  Stackelberg game 
  & \vspace{-1em} \begin{itemize}[leftmargin=*, nosep, topsep=0pt, partopsep=0pt] \item Multi-party energy management problem of a microgrid which includes prosumers with \ac{ESS} and \ac{PEV} charging stations is addressed, respectively energy management strategies were determined.  \end{itemize} \vspace{-1em}  & \cite{erol2022stackelberg}\\
  & \vspace{-1em} \begin{itemize}[leftmargin=*, nosep, topsep=0pt, partopsep=0pt] \item Proposed an energy sharing scheme considering Regioin-to-Region(R2R) and demand resopnse. \item  The practical case study effectively demonstrates improving local energy consumption and the areas' economic interests.  \end{itemize} \vspace{-1em}  & \cite{10187671}\\
 & \vspace{-1em} \begin{itemize}[leftmargin=*, nosep, topsep=0pt, partopsep=0pt] \item The platform is modeled in such a way that retailers act as leaders,  CS and \ac{EV}s act as the followers to maximize the profits and minimize the energy cost respectively. \item In addition penalty factor is introduced to enhance social welfare.
   \end{itemize} \vspace{-1em}   & \cite{adil2021energy}\\
 & \vspace{-1em} \begin{itemize}[leftmargin=*, nosep, topsep=0pt, partopsep=0pt] \item Benefits for both producers and consumers are achieved. Furthermore, as the number of prosumers in the virtual microgrid rises and CO$_2$ emissions fall.  \end{itemize} \vspace{-1em}   & \cite{anoh2019energy}\\
 &  \vspace{-1em} \begin{itemize}[leftmargin=*, nosep, topsep=0pt, partopsep=0pt] \item  Distribution system operator problem is expressed as a mixed-integer second-order cone programming (MISOCP) model.\item The effectiveness of the proposed model is studied on a 4-microgrid connected to a 33-bus and 123-bus distribution test system.  \end{itemize} \vspace{-1em} & \cite{9235517}\\
  \hline
  \end{tabular}
\end{table*}

\subsection{Blockchain Technology}
\begin{figure}
    \centering
    \includegraphics[width=1\linewidth]{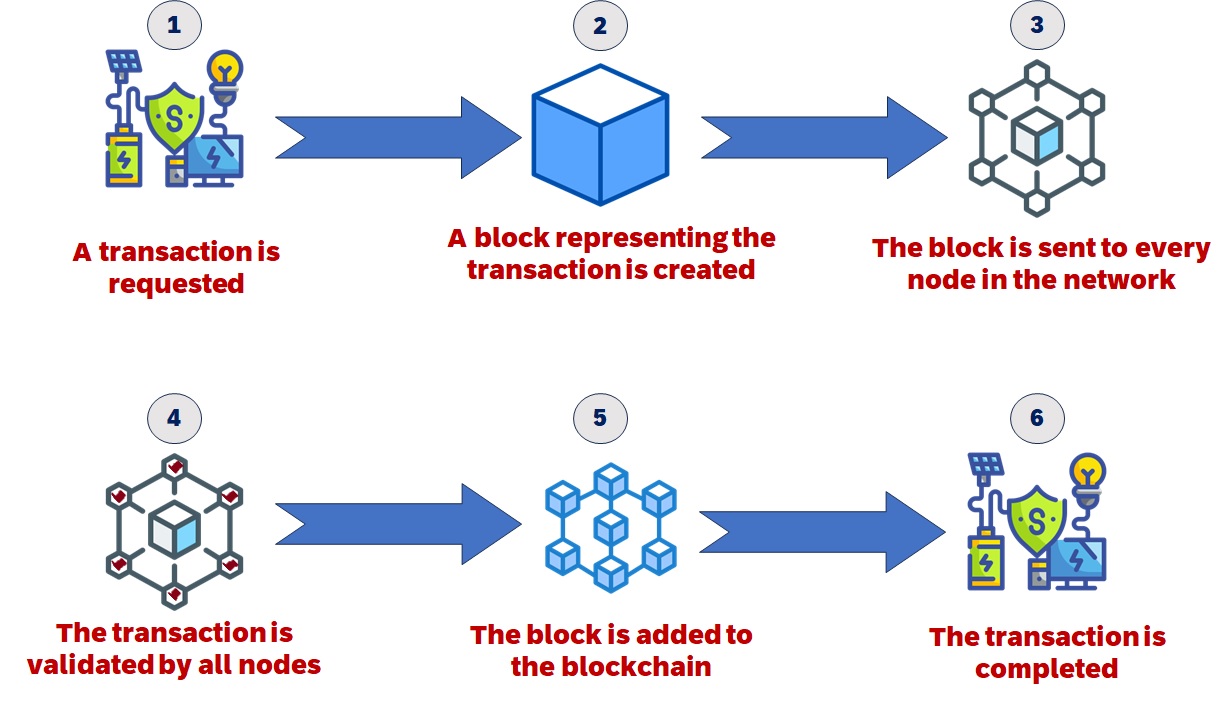}
    \caption{Blockchain Technology}
    \label{fig:BCT}
\end{figure}

Blockchains are dispersed digital ledgers of transactions that are cryptographically signed and organized into blocks. The process that occurs in the blockchain technology is depicted in the Fig \ref{fig:BCT}. After verification and consensus-building, every block is cryptographically connected to the one before it, rendering any tampering obvious. Tamper resistance arises from the difficulty of altering older blocks when new ones are introduced. Within the network, new blocks are replicated across copies of the ledger, and any conflicts are automatically resolved by applying pre-established criteria \cite{yaga2019blockchain}, \cite{7998367}. In addition to the architecture, typical consensus algorithms are presented in \cite{8029379}. The impact of blockchain technology with use cases is studied in \cite{efanov2018all}. Due to the versatility of blockchain technology, its implementation and challenges are discussed in \cite{andoni2019blockchain} and in addition, different start-ups that have emerged in the energy sector, \ac{P2P}  in \ac{EV} charging, and possible applications are presented in \cite{thukral2021emergence}. Table \ref{tab:Blockchain-consortium} and  Table \ref{tab:Blockchain_ethereum} gives insights into some articles related to blockchain technology implementation in \ac{P2P} trading.

\begin{table*}
    \centering
    \setlength\extrarowheight{1mm}
  \caption{Machine Learning Algorithms}
  \label{tab:ML}
    \begin{tabular}{|p{2cm}|p{13cm}|p{1.2cm}|}
    \hline
  \textbf{Algorithm} & \textbf{Contribution} & \textbf{Reference} \\
  \hline
Deep Reinforcement Learning 
  & \vspace{-1em} \begin{itemize}[leftmargin=*, nosep, topsep=0pt, partopsep=0pt] \item A large number of heterogenous households are classified as four clusters, and case studies exhibited a strong generalization capability. 
  \end{itemize} \vspace{-1em}  & \cite{qiu2021scalable}\\
& \vspace{-1em} \begin{itemize}[leftmargin=*, nosep, topsep=0pt, partopsep=0pt] \item Significant reduction in  operation cost for each microgrid. \end{itemize} \vspace{-1em}  & \cite{9596598}\\
& \vspace{-1em} \begin{itemize}[leftmargin=*, nosep, topsep=0pt, partopsep=0pt] \item A use case of multi-agent deep deterministic policy gradients (MADDPG) with energy arbitrage and RES integration. \item Evaluated the use of a single-agent global controller over multiple distributed agents to control the microgrid components. \end{itemize} \vspace{-1em}  & \cite{harrold2022renewable}\\
 & \vspace{-1em} \begin{itemize}[leftmargin=*, nosep, topsep=0pt, partopsep=0pt] \item  Multi-agent model for double auction regional microgrid and a MARL framework for interactive learning is designed. \item The proposed MARL framework improved the operational performance of the microgrid and enabled benefits to participants. \end{itemize} \vspace{-1em} & \cite{fang2021multi}\\
& \vspace{-1em} \begin{itemize}[leftmargin=*, nosep, topsep=0pt, partopsep=0pt] \item MA-POMDP is used for describing the agents under communication failures. BA-DRL is applied to estimate the Q-values, and simulation results show the method's robustness. \end{itemize} \vspace{-1em}   & \cite{9631955}\\
&  \vspace{-1em} \begin{itemize}[leftmargin=*, nosep, topsep=0pt, partopsep=0pt] \item  The prioritized experience replay and multi-actor attention-critic approaches' advantageous features are combined in this model.\item  Case studies of large-scale real-time operations show satisfactory results. \end{itemize} \vspace{-1em}   & \cite{9705504}\\
&  \vspace{-1em} \begin{itemize}[leftmargin=*, nosep, topsep=0pt, partopsep=0pt] \item The transaction process is constructed as MDP with a deep reinforcement learning algorithm. \item An indirect customer-to-customer multi-energy transaction is modeled, followed by a multi-time scale hybrid trading mechanism. \end{itemize} \vspace{-1em} & \cite{9681381}\\
& \vspace{-1em} \begin{itemize}[leftmargin=*, nosep, topsep=0pt, partopsep=0pt] \item  Continuous double auction market is modeled as decentralized partially observed \ac{MDP} and then novel multi-agent \ac{RL} method is used. \item For stabilizing each prosumer agent training performance mean-field approximation is used. \item Case studies are conducted on the real-time scenario of 100 prosumers, and the proposed method is applied to show the economic benefits.  \end{itemize} \vspace{-1em} & \cite{9931995}\\
\hline
Neural network  
& \vspace{-1em} \begin{itemize}[leftmargin=*, nosep, topsep=0pt, partopsep=0pt] \item The model is trained by BPNN with time of day and temperature as inputs, and modified LSTM deep learning is used to correct the trained network.\item  Fast economics scheduling is done by a modified consistency algorithm at each microgrid layer.\item   For multiple microgrids, an adaptive Tchebycheff-based MPEA/D is used.  \end{itemize} \vspace{-1em}    & \cite{tan2020multi}\\
 & \vspace{-1em} \begin{itemize}[leftmargin=*, nosep, topsep=0pt, partopsep=0pt] \item A hybrid algorithm is designed that uses reinforcement learning and a feedforward neural network. \item Minimization of network latency, processing time, and packet error and proof-of-work validation for successful transactions are executed. \end{itemize} \vspace{-1em}  & \cite{shukla2023network}\\
\hline
Extreme learning machines  
& \vspace{-1em} \begin{itemize}[leftmargin=*, nosep, topsep=0pt, partopsep=0pt] \item The model consists of four features, namely data cleaning, feature selection, prediction, and parameter optimization. \item MOGOA is used for feature selection, and DELM is applied for forecasting the load requirements. \item The model is evaluated using the UK smart meter dataset. \end{itemize} \vspace{-1em}  & \cite{varghese2022optimal}\\
\hline
Reinforcement learning 
& \vspace{-1em} \begin{itemize}[leftmargin=*, nosep, topsep=0pt, partopsep=0pt] \item Designed a local energy market (LEM), proposed ETS that manages trades in LEM then CES  is applied to the LEM and controlled by ETS. \item The proposed two-phase model has achieved maximum profits. \end{itemize} \vspace{-1em}   & \cite{zang2021reinforcement}\\
& \vspace{-1em} \begin{itemize}[leftmargin=*, nosep, topsep=0pt, partopsep=0pt] \item  The proposed SynergyChain is developed in Python and tested using Ethereum test nets. \item  Integrating the reinforcement learning module improved the overall system performance. \end{itemize} \vspace{-1em}  & \cite{9305283}\\
& \vspace{-1em} \begin{itemize}[leftmargin=*, nosep, topsep=0pt, partopsep=0pt] \item Designed MDP and a reinforcement learning algorithm was applied that enhance the performance and numerical analysis is conducted to evaluate the performance.  \end{itemize} \vspace{-1em}   & \cite{8661902}\\
\hline
Deep learning 
& \vspace{-1em} \begin{itemize}[leftmargin=*, nosep, topsep=0pt, partopsep=0pt] \item The proposed framework generates blocks using short signature and hash functions. \item An intrusion detection system(IDS) is designed to detect network attacks and fraudulent transactions. \item The proposed IDS performance is studied on three different CICIDS2017 datasets.  \end{itemize} \vspace{-1em}  & \cite{8758147}\\
  \hline
  \end{tabular}
  \end{table*}

\begin{table*}
    \centering
    \setlength\extrarowheight{2mm}
     \caption{Optimization \& Other Models}
  \label{tab:AM}
    \begin{tabular}{|p{2cm}|p{4cm}|p{9cm}|p{1.2cm}|}
    \hline
  \textbf{Type} & \textbf{Methodology } & \textbf{Contribution} & \textbf{Reference} \\
  \hline
Model &  Cascaded model predictive control
optimization framework. & \vspace{-1em} \begin{itemize}[leftmargin=*, nosep, topsep=0pt, partopsep=0pt] \item The first MPC scheme is called the energy layer, the second MPC scheme called the transport layer is designed.\item The model is validated by conducting a case study in Tokyo. \end{itemize} \vspace{-1em} & \cite{8909735}\\
&  a novel P2P energy trading model.  & \vspace{-1em} \begin{itemize}[leftmargin=*, nosep, topsep=0pt, partopsep=0pt] \item Based on several critical success criteria, such as the market structure, trading mechanism, physical and virtual infrastructure, policy and governance, and social aspects, a model for peer-to-peer energy trading is put into place for Malaysia.\item The model is to study the implications of \ac{P2P} trading in Malaysia \end{itemize} \vspace{-1em} & \cite{9739692}\\
&  Normalized P2P (NP2P) energy trading scheme is modeled as Knapsack Problem (KP) and solved using Greedy and Simulated Annealing (SA) algorithms.  & \vspace{-1em} \begin{itemize}[leftmargin=*, nosep, topsep=0pt, partopsep=0pt] \item The suggested \ac{P2P} energy trading scheme's effectiveness is tested in three distinct scenarios.\item To confirm the efficacy of the solution, the outcomes of several assessed cases for the suggested NP2P scheme and ECP algorithms are compared.\end{itemize} \vspace{-1em} & \cite{9598827}\\
&  A novel model-based, multi-agent asynchronous advantage actor-centralized-critic with communication (MB-A3C3) approach.  & \vspace{-1em} \begin{itemize}[leftmargin=*, nosep, topsep=0pt, partopsep=0pt] \item The model uses a large hourly 20122013 dataset of 300 Sydney residences with rooftop solar systems in NSW, Australia.\item The MB-A3C3 reinforcement learning system lowers community energy expenses by 17 \% for 300 families more than MADDPG and A3C3.\end{itemize} \vspace{-1em} & \cite{9963562}\\
&  a data-driven distributionally robust co-optimization model.  & \vspace{-1em} \begin{itemize}[leftmargin=*, nosep, topsep=0pt, partopsep=0pt] \item Studies on four interconnected microgrids confirm the benefits of the proposed P2P energy trading network and show that the DRO model handles uncertainty better than the robust optimization (RO) and stochastic programming (SP) models.\end{itemize} \vspace{-1em} & \cite{9477425}\\
& A novel integrated transaction and operation model.  & \vspace{-1em} \begin{itemize}[leftmargin=*, nosep, topsep=0pt, partopsep=0pt] \item Numerical case studies using IEEE 13-bus distribution network prove that the suggested technique helps prosumers trade energy for distribution network operational safeguard issues and reduce consumption costs.\end{itemize} \vspace{-1em} & \cite{9733902}\\ 
&  a bidding-based peer-to-peer (P2P) energy transaction optimization model.  & \vspace{-1em} \begin{itemize}[leftmargin=*, nosep, topsep=0pt, partopsep=0pt] \item Objective function weighted social-welfare terms give different community operations reflecting energy trading operator (community operator) goal. \item The case study found that prosumers and communities benefit from energy transactions and green energy choice reduces CO2 emissions. \end{itemize} \vspace{-1em} & \cite{9361558}\\

\hline
Optimization strategies & A cooperative game theory technique based on \ac{PSO}. & \vspace{-1em} \begin{itemize}[leftmargin=*, nosep, topsep=0pt, partopsep=0pt] \item  A typical 3-layered networked microgrid is designed then MOP is performed.\item A two-stage control framework is designed for P2P, P2G energy trading.\item  An Australian test case is considered for analysis. \end{itemize} \vspace{-1em} & \cite{ali2021multi} \\
&  Energy Balance service provider (EBSP) considering market elasticity  & \vspace{-1em} \begin{itemize}[leftmargin=*, nosep, topsep=0pt, partopsep=0pt] \item For a community microgrid, a market equilibrium model is designed.\item To achieve maximum profitability for EBSP, an optimized pricing and trade strategy has been put in place.\end{itemize} \vspace{-1em} & \cite{wang2021peer} \\
& Slime-mould inspired optimisation method. & \vspace{-1em} \begin{itemize}[leftmargin=*, nosep, topsep=0pt, partopsep=0pt] \item An optimal energy routing path is designed with capacity constraints taken into account. \item The proposed method is flexible with a large no of peers. \end{itemize} \vspace{-1em} & \cite{9097579}\\
& Parametric optimization & \vspace{-1em} \begin{itemize}[leftmargin=*, nosep, topsep=0pt, partopsep=0pt] \item An energy management model for each peer is proposed to minimize the cost. \item The influence of multi-energy coupling on \ac{P2P} transaction is studied. \end{itemize} \vspace{-1em} & \cite{zhang2022parametric}\\
& Energy Trading Distributed Alternating Direction Method of Multipliers (ETD-ADMM). & \vspace{-1em} \begin{itemize}[leftmargin=*, nosep, topsep=0pt, partopsep=0pt] \item  Along with economic benefits of \ac{P2P} trading, Quantitative and qualitative comparison within distributed algorithms are demonstrated.\item It is implemented in test cases that demonstrate the effectiveness and feasibility of the model. \end{itemize} \vspace{-1em} & \cite{umer2021novel}\\
\hline
\end{tabular}
\end{table*}

\subsection{Game Theory Approaches}
\begin{figure}
    \centering
    \includegraphics[width=1\linewidth]{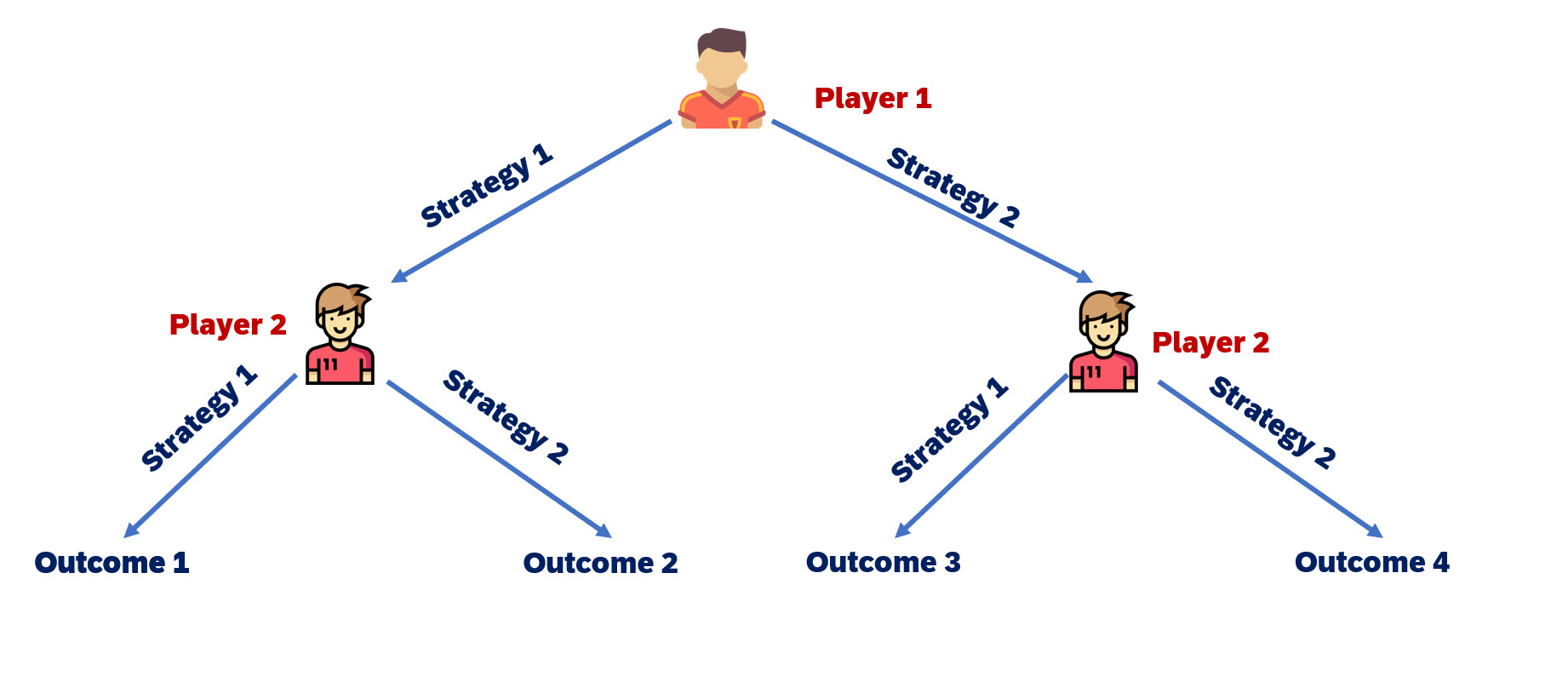}
    \caption{Game Theory}
    \label{fig:GT}
\end{figure}

In the context of economic agents, game theory analyzes how their interactions result in outcomes that are aligned with their preferences or utilities, even when none of the agents might have intentionally sought those specific outcomes. Fig \ref{fig:GT} gives the basic idea of the game theory approach in which players participate to obtain maximum benefits. Each player's strategy is based on the other player's strategies. The book \cite{osborne2004introduction} explains the fundamental concepts of game theory and demonstrates how to apply them to comprehend biological, social, political, and economic phenomena. \ac{P2P} energy trading emerged as an energy management mechanism in the electricity market that inclusively coordinates prosumers and the grid, although this needs a strong mathematical and signal principle for decision making, so the game theory approach provides this decision-making capability. Application of game theory to solve power systems problems is given in \cite{abapour2020game}. Meanwhile, Wayes Tushar et al.\cite{8398582} gave an overview of the use of the game theoretic approach in \ac{P2P} energy trading. Game theory approaches are categorized as cooperative and non-cooperative games, symmetric and asymmetric, and simultaneous and sequential games, as given in Fig. \ref{fig:ETMCA}. Table \ref{tab:GT} illustrates the game theory approaches and methodologies. 

\subsection{Machine Learning}
\begin{figure}
     \centering
     \includegraphics[width=1\linewidth]{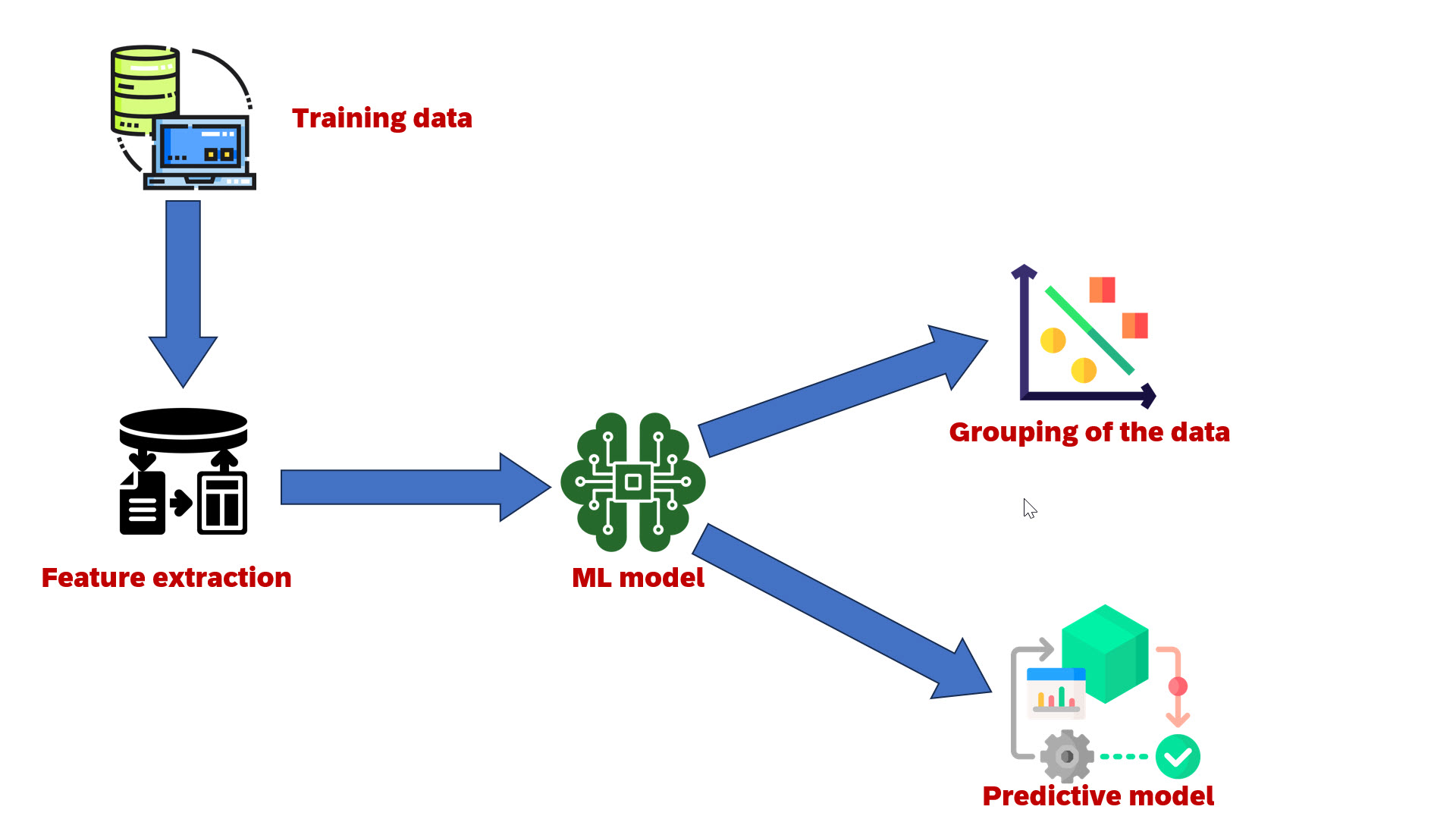}
     \caption{Machine Learning Approach}
     \label{fig:ML}
 \end{figure}

 A branch of artificial intelligence, \ac{ML} is based on the idea that the system learns from the data, identifies patterns, and makes decisions with minimal human intervention. The basic Fig \ref{fig:ML} explains the machine learning principle and its working. ac{ML} basics were discussed in \cite{wang2016machine}, \cite{alpaydin2016machine}, whereas various machine algorithms were covered in \cite{mahesh2020machine} and application of \ac{ML} in power systems analytics is studied by Seyed Mahdi Miraftabzadeh et al.\cite{8783340} and application in \ac{SG} and energy internet studied in  \cite{cheng2019new}, Yize Chen et al. \cite{8587547} gave their insights on the vulnerabilities regarding the security of \ac{ML} in power. Power systems fault diagnosis applications are studied in \cite{vaish2021machine}, whereas power systems resilience is studied in \cite{xie2020review}. \ac{ML} algorithms are also used in \ac{P2P} energy trading and Table \ref{tab:ML}  gives complete overview of the application \ac{ML} algorithms.

\subsection{Optimization Stratergies \& Other Models}
 Optimization is a method in which an objective function is mathematically modeled and solved to predict the behavior of a process or system. To minimize or maximize the objective function the selection of an optimization technique is important. In \cite{alfares2023introduction} fundamentals of optimization concepts,  various optimization models, and solutions were discussed. Meanwhile in \cite{emmerich2018tutorial}, mathematical modeling of multi-objective optimization is performed and evolutionary or bio-inspired methods are discussed. The optimization approaches in power systems are either applied to minimize losses or to maximize profits and \cite{momoh2017electric} focuses on the implementation of optimization techniques in power system problems. The use of \ac{P2P} energy trading in particular helps to maximize the financial gains for prosumers, consumers, and \ac{EVCS}. It also keeps the balance between the production and consumption of energy, including losses. The \ac{P2P} electricity trading platforms were developed utilizing constrained optimization techniques, including mixed integer linear programming, non-linear optimization, the alternating directing method of multipliers, and linear programming. There are several optimization techniques reported in the literature\cite{venter2010review},\cite{hobbs1995optimization}. Table\ref{tab:AM} gives an overview of the models and optimization techniques researchers use.
\section{Challenges and Future Scope}
The integration of energy sharing the electrical distribution system between microgrids and \ac{EVCI} poses a variety of challenges to both physical and virtual layers. Infrastructure compatibility issues present physical layer obstacles that need to be upgraded to guarantee smooth integration between \ac{EVCI}, microgrids, and the current grid infrastructure. Effective balancing methods are necessary to address capacity challenges posed by grid congestion in locations with a large concentration of \ac{EVCI}. It is a big job to coordinate the integration of energy storage devices inside the microgrid to control variations in the demand for \ac{EV} charging and the production of renewable energy. Continuous energy-sharing activities depend on the \ac{EVCI} and microgrid hardware being reliable, which only be achieved by routine maintenance and monitoring. Furthermore, as clear policies supporting and motivating energy-sharing projects are critical for success, overcoming regulatory hurdles relating to grid connectivity, participation in the energy market, and tariff structures is necessary.

In addition, Cybersecurity becomes a major issue at the virtual layer, requiring strong defenses against cyberattacks on control systems and communication networks. Another challenge is to harmonize communication protocols and standards since interoperability problems prevent \ac{EVCI} and microgrid components from exchanging data seamlessly. Furthermore, to reduce concerns over information sharing, it is essential to establish clear standards regarding data privacy and ownership. The virtual layer develops increasingly advanced with smart grid management, enabling intelligent control systems that optimize energy flows by taking demand-response mechanisms and grid stability into consideration
\subsection{Challenges for \ac{EVCI} Sharing}
\begin{figure}
    \centering
    \includegraphics[width=1\linewidth]{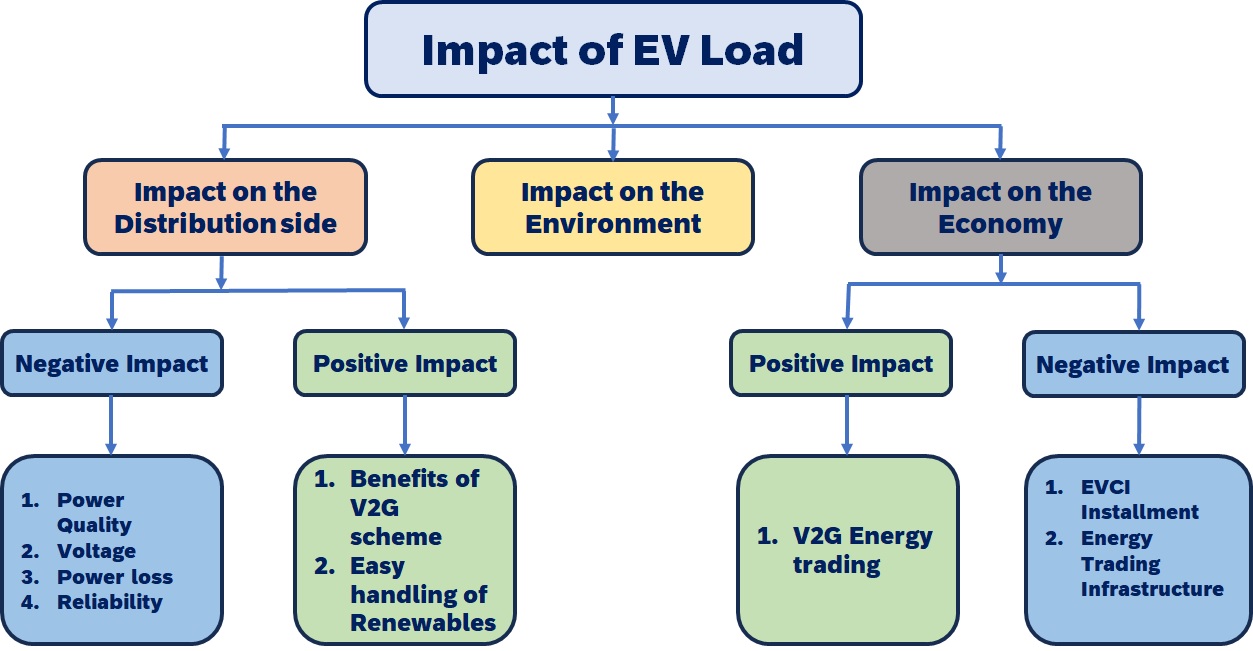}
     \caption{Impact of \ac{EV} }
     \label{fig:Impact-ev}
\end{figure}

\ac{EV}s are playing an important role in reducing the carbon footprint and \ac{GHG}. The transportation industry has been transformed by \ac{EV}, and all worldwide statistics indicate that the number of \ac{EV}s will increase significantly in the future, posing several challenges. This includes \ac{EVCI} and corresponding policies and \ac{EV} charging impacts; these both are interrelated. The first thing is that \ac{EVCI} is entirely dependent on the kind of charging station. This requires significant investments and a great deal of technological advancement. This poses a greater challenge to governments across the nations to invest in it. \cite{metais2022too}  examined charging infrastructure deployment models and covered related technical, financial, and user behavior-related concerns. In addition to providing a comprehensive review of the literature, \cite{ashfaq2021assessment} looks at the effects of EVs on the electric grid and the most recent deployment and difficult problems with implementing EV infrastructure, charging power levels, and various charging techniques. Fig. \ref{fig:Impact-ev} illustrates the impact of the \ac{EVCS} n particular \ac{EVCS}s on the distribution side and in particular on the environment. 

\subsubsection{Challenges for Electrical Distribution Systems}

The challenges that impact the \ac{EDS} due to \ac{EVCS} energy sharing are power quality issues, voltage, power loss, and reliability. The impact of \ac{EVCS} on \ac{EDS} was studied in \cite{6905855}, and \cite{7264982}. Firstly, the widespread adoption of the \ac{EV}s will lead to an increase in the energy demand, particularly during the peak hours. This increasing demand imposes a burden on the distribution transformers, distribution feeders, and in particular, distribution substations, leading to the destabilization of the distribution systems. It overloads the grid, which results in power quality issues like voltage fluctuations.\cite{wang2018impact}, \cite{rahman2022comprehensive}, and \cite{bass2013impacts} gave their insights on power quality issues with the integration of \ac{EV} on the power grid in particular low voltage \ac{EDS}. Secondly, higher concentrations of the \ac{EVCS}s will pose challenges to the voltage regulations due to the fast charging during the peak hours, which results in distribution system instability. In addition, the voltage fluctuations will affect the performance of the equipment and will lead to an increase in the distribution power losses. Lastly, the concentration of charging stations in urban locations where the demand is high will lead to grid congestion. This will result in slow charging and reduced reliability; in addition, the consumers will have to pay higher costs for electricity usage. 

\subsubsection{Economic Challenges}
Infrastructure upgradation and \ac{EV} trading infrastructure development are two critical challenges, and, importantly, it require huge investments.  Firstly, the \ac{EDS} upgradation should be completed to cope with the energy demand. Meanwhile, the \ac{EVCI} is meticulously planned as it is based on the type of charging station and the behavior patterns of the \ac{EV} user to avoid additional costs\cite{metais2022too}.  However, the electricity trading system currently in place has a wide range of challenges including energy price volatility, demand variability, capacity, and temporal uncertainties, congestion at charge stations, storage systems capacities, security and privacy protection, voltage and frequency regulation, communication overheads, pricing strategies etc., \ac{EV} trade needs to be taken into account as well\cite{adil2021energy}. 

\subsection{Challenges for Microgrid}

\subsubsection{Challenges for Electrical Distribution systems}

\begin{figure}
    \centering
    \includegraphics[width=1\linewidth]{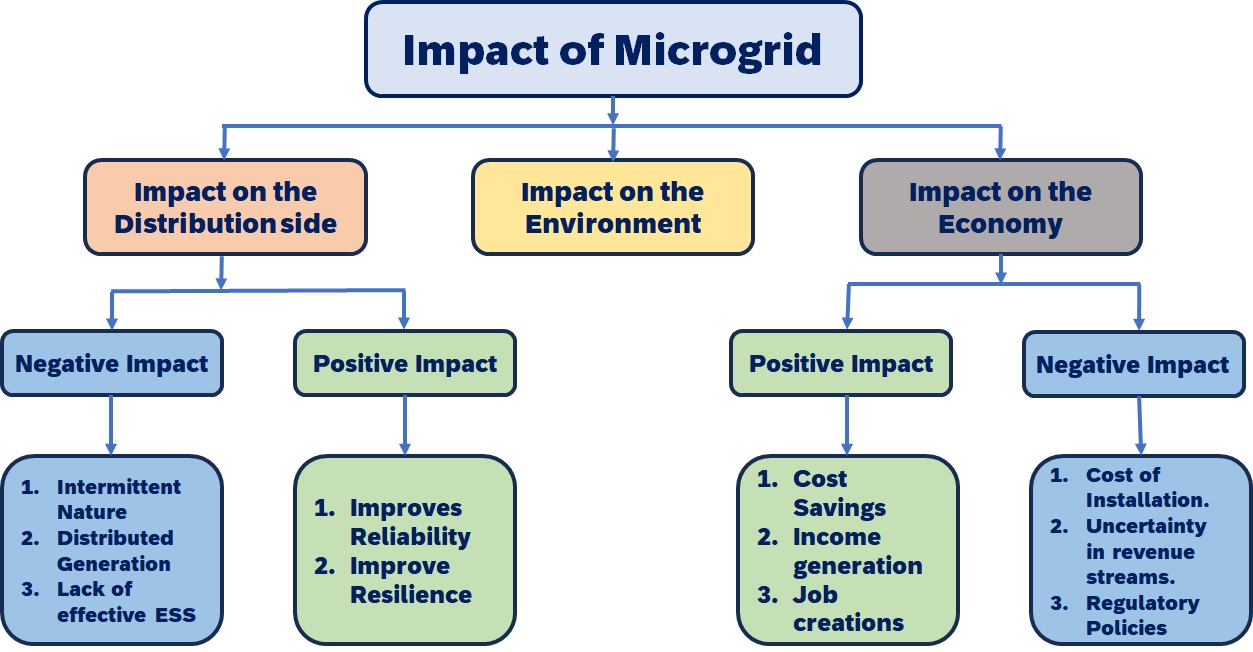}
    \caption{Impact of Microgrid sharing}
     \label{fig:MG_shar}
\end{figure}

The challenges of microgrid integration with \ac{EDS} are as follows: intermittent nature, distributed generation, lack of effective \ac{ESS}s, and interoperability, as shown in Fig. \ref{fig:MG_shar}. Due to the intermittent nature of the microgrid generation, integrating with the main grid is complex. It is a challenging task to coordinate with proper protocols for continuous interaction and control. As the \ac{DER} are distributed, this requires regulatory frameworks and policies to avoid uncertainties for all the stakeholders. \ac{ESS} plays a key role in addressing problems related to power quality, reliability issues, stability, and effective energy management\cite{nazaripouya2019nergy}. But \ac{ESS} also has numerous issues in its integration, saftey, \ac{SoC}, \ac{SoD}, life span, capacity, reliability and cost \cite{choudhury2022review}; so effective \ac{ESS}s are required to address the above issues. Lastly, the major challenge faced by microgrids is interoperability.  It is difficult for full communication to be achieved as there are no common protocols or standards among the various microgrid components and control systems. To ensure interoperability, the industry must focus on the development and adoption of standardized interfaces and protocols.

\subsubsection{Economic Challenges}
The economic challenges include the cost of installations through financial support from the governments, uncertainties in revenue generation, and lack of regulatory framework and policies. Investing in microgrids is still expensive. Without some form of financial assistance, several of its components such as fuel cells, photovoltaic (PV), and power storage—are not yet commercially feasible \cite{bellido2018barriers}.

\subsection{Cybersecurity Challenges}
The incorporation of microgrids and \ac{EVCI} for energy sharing into electrical distribution systems presents several cybersecurity challenges. It is critical to ensure the security of these systems to avert disruptions, unauthorized access, and data manipulation. Certain cybersecurity challenges are unique to the integration of microgrids and electric vehicle infrastructure Cyber vulnerability analysis conducted on \ac{EVCS} through simulation that assists in predicting the challenges corresponding to cybersecurity is discussed in \cite{9490069}. The cybersecurity challenges in microgrids were discussed in \cite{canaan2020microgrid}:

\subsubsection{Access Irregularity and Verification}
The presence of weaknesses in authentication mechanisms or insufficient access controls could potentially result in unauthorized entry into \ac{EV} infrastructure and microgrid systems. Data breaches, manipulation of energy transactions, and possible disruptions to \ac{EVCS} and microgrid operations may result from unauthorized access\cite{novak2023network}.

\subsubsection{Confidentiality and Data Integrity}
The integrity and confidentiality of data exchanged among \ac{EV}s, charging stations and microgrid components must be strictly maintained. Unauthorized access to sensitive information or data tampering compromises the security and dependability of energy transactions and system operations. Examples of how to create and detect/mitigate False Data Injection (FDI) threats in smart microgrids are given in \cite{nejabatkhah2020cyber} along with examples of current global cyber-security initiatives as well as essential smart grid cyber-security guidelines are provided.

\subsubsection{Ensuring Communication Security}
One potential challenge is the vulnerability of insecure communication channels connecting \ac{EV}s, charging stations, and microgrid components to surveillance and man-in-the-middle attacks. Unauthorized control over \ac{EV} charging, data interception, and possible disruptions to microgrid operations may result from compromised communication. Active distribution network communication technologies, together with their applications and communication standards, are examined and reviewed in \cite{9987475}.

\subsubsection{Ensuring Endpoint Security}
Endpoints that lack sufficient security measures, including microgrid controllers and \ac{EVCS}, could potentially be susceptible to malware, ransomware, and other forms of cyber threats. Energy sharing disruptions, unauthorized control over charging stations, and potential safety risks may result from compromised endpoints\cite{novak2023network}.

\subsubsection{Supply Chain Protection}It is of the utmost importance to safeguard components and systems procured from diverse suppliers against supply chain breaches. The potential introduction of vulnerabilities by compromised components could result in cybersecurity incidents and cause disruptions to the operations of electric vehicles and microgrids. These disruptions to supply chains and the functioning of these systems' embedded hardware devices are among the threats\cite{moghadasi2022trust}.

\subsubsection{Software and firmware susceptibilities}
Significant risk is posed by unpatched or insecure software and firmware in electric vehicles, charging stations, and microgrid controllers. Vulnerabilities that are exploited may result in unauthorized access, control, or manipulation of critical system functions\cite{upadhyay2020scada}. Data tampering causes inaccurate billing or even causes physical harm and illegal access results in stealing of the personal information or control charges \cite{xuefeng2022risks}. 

\subsubsection{Denial of service(DoS) attacks} DoS attacks that are directed at the \ac{EVCI} or microgrid components have the potential to disrupt energy transactions and overall system operations. Service interruptions may cause financial losses, inconvenience for \ac{EV} users, and even pose safety risks\cite{9233930}.

\subsubsection{cyber eavesdropping}
Insiders, such as contractors and workers, have the potential to intentionally or unintentionally jeopardize the security of microgrids and \ac{EVCI}. Insider threats result in unauthorized access, data breaches, and disruptions to energy-sharing operations. A thorough analysis of Man-in-the-Middle Attacks was covered in \cite{novak2023network} along with the countermeasures. 

\subsubsection{Regulatory Conformity} Organizations may face difficulties in ensuring compliance with cybersecurity regulations and standards in the energy sector. Consequences of noncompliance include legal implications, harm to one's reputation, and heightened susceptibility to cyber threats.

\subsubsection{Security Inadequateness}Insufficient knowledge and instruction regarding cybersecurity among users, administrators, and stakeholders could result in substandard practices. Instances of human error, including susceptibility to phishing attacks and disregard for security best practices, have the potential to introduce vulnerabilities into the system. It is necessary to assess the knowledge gap between cyber security education and industry needs due to the alarming rise in cyber incidents and severe skills scarcity and analysis of cybersecurity knowledge gaps is studied in \cite{catal2023analysis}.

To effectively tackle these cybersecurity challenges, a holistic strategy is necessary, which includes implementing strong security protocols, conducting routine vulnerability assessments, educating personnel on best practices, and adhering to industry standards. Industry stakeholders, regulatory bodies, and cybersecurity specialists must work in concert to establish an energy-sharing ecosystem that is both resilient and secure.

\section{Conclusions}
Over a few decades, the \ac{EDS} has seen significant changes, mainly due to the installment of \ac{DER}, and microgrids resulting in the utilization of renewables. Moreover, the adoption of the \ac{EV}s further restructures the \ac{EDS}. This will play a pivotal role in changing the energy landscape, and the decentralization of the present-day grid is the key indicator for the change. Notably, the prosumers, who produce and consume energy, are the key to the decentralized grid, consequently leading to \ac{P2P} energy sharing. As \ac{P2P} energy sharing poses many economic and societal challenges, it requires comprehensive research, and it opens a key initiative for \ac{P2P} energy trading. So, many researchers have dedicated their efforts to promoting the transition towards the \ac{P2P} energy markets aims to integrate the prosumers into the distribution systems, leading to wider utilization of the renewables.

The primary contribution of this paper lies in offering a comprehensive overview of the energy-sharing concept and sharing models with case studies. In addition, a detailed overview of the infrastructure needs for energy-sharing in the physical and virtual layers was also presented by discussing the components of the physical and virtual layers. The technical concept behind \ac{P2P} energy trading is discussed, and the implementation of \ac{P2P} energy trading is presented as market clearing algorithms. By rigorously reviewing the existing literature, this paper organizes the market-clearing algorithms into five categories, namely double auction theory, blockchain technology, game theory, Machine learning, and optimization strategies \& other models. The findings have the potential to guide policymakers, researchers, and industry stakeholders toward a sustainable, resilient, and efficient energy landscape.

This review study presents social, and economic aspects for the future,  leading to multidisciplinary research. This paper highlights interdisciplinary opportunities such as machine learning and game theory in diverse disciplines including psychology and economics for energy trading. Additionally, the recommendations encourage the involvement of the stakeholders in the electricity market. In essence,  this paper presents various aspects of energy sharing among peers that require substantial research efforts for a successful transition to real-world applications.

\bibliographystyle{IEEEtran}
\bibliography{main}

\begin{thebibliography}{100}
\bibitem{Sonnen}
``We are the connection to a sustainable energy future.'' [online]. Available: sonnengroup.com/sonnencommunity.
\end{thebibliography}

\end{document}